\documentclass[aps, prd, 12pt, showpacs, floatfix, twocolumn, nofootinbib, reprint, preprintnumbers, longbibliography]{revtex4-2}

\usepackage{multirow}
\usepackage{booktabs}  
\usepackage{makecell}  

\usepackage{amsmath, hyperref}
\usepackage{dcolumn}
\usepackage[dvipsnames]{xcolor}
\usepackage{graphicx}
\usepackage{subcaption}
\usepackage{graphicx}
\usepackage[autostyle]{csquotes}
\usepackage[final]{microtype}
\hypersetup{
  colorlinks=true,
  citecolor=red,
  linkcolor=magenta,
  urlcolor=NavyBlue,
  allbordercolors={0 0 0},
  pdfborderstyle={/S/U/W 1}
}

\usepackage{orcidlink}

\def \Hz	    {\mathrm{Hz}}

\def \param     {\vec \Lambda}

\def \IITGn     {Indian Institute of Technology Gandhinagar, Gujarat 382055, India.\vspace*{4pt}}

\begin{document}

\title{Pinpointing coalescing binary neutron star sources \\with the IGWN, including LIGO-Aundha}

\author{\textsc{Sachin~R.~Shukla}\orcidlink{0009-0009-8046-6161}}
\email{sachins@alumni.iitgn.ac.in }
\affiliation{\IITGn}

\author{\textsc{Lalit~Pathak}\orcidlink{0000-0002-9523-7945}}
\email{lalit.pathak@iitgn.ac.in}
\affiliation{\IITGn}

\author{\textsc{Anand~S.~Sengupta}\orcidlink{0000-0002-3212-0475}\vspace*{5pt}} 
\email{asengupta@iitgn.ac.in}
\affiliation{\IITGn}

\begin{abstract}
LIGO-Aundha (A1), the Indian gravitational wave detector, is expected to join the International Gravitational-Wave Observatory Network (IGWN) and begin operations in the early 2030s. We study the impact of this additional detector on the accuracy of determining the direction of incoming transient signals from coalescing binary neutron star (BNS) sources with moderately high signal-to-noise ratios. 
It is conceivable that A1's sensitivity, effective bandwidth, and duty cycle will improve incrementally through multiple detector commissioning rounds to achieve the desired `LIGO-A+' design sensitivity. For this purpose, we examine A1 under two distinct noise power spectral densities. One mirrors the conditions during the fourth science run (O4) of the LIGO Hanford and Livingston detectors, simulating an early commissioning stage, while the other represents the A+ design sensitivity. We consider various duty cycles of A1 at the sensitivities mentioned above for a comprehensive analysis. 
We show that even at the O4 sensitivity with a modest $20\%$ duty cycle, A1's addition to the IGWN leads to a $15\%$ reduction in median sky-localization errors ($\Delta \Omega_{90\%}$) to $5.6$~sq.~deg. At its design sensitivity and $80\%$ duty cycle, this error shrinks further to $2.4$~sq.~deg, with 84\% sources localized within a nominal error box of $10$~sq.~deg. This remarkable level of accuracy in pinpointing sources will have a positive impact on Gravitational Wave (GW) astronomy and cosmology. Even in the worst-case scenario, where signals are sub-threshold in A1, we demonstrate its critical role in reducing the localization uncertainties of the BNS source. 
Our results are obtained from a large Bayesian parameter estimation study using simulated signals injected in a heterogeneous network of detectors using the recently developed meshfree approximation aided rapid Bayesian inference pipeline. We consider a seismic cut-off frequency of 10 Hz for all the detectors.
We also present hypothetical improvements in sky localization for a few Gravitational-Wave Transient Catalog (GWTC)-like events injected in real data after including a hypothetical A1 detector to the sub-network in which such events were originally detected. We also demonstrate that A1's inclusion could resolve the degeneracy between the luminosity distance and inclination angle parameters, even in scenarios where A1 does not directly contribute to improving the network signal-to-noise ratio for the event.
\end{abstract}
\maketitle

\section{Introduction}
\label{sec:intro}

\label{subsec: GW170817 and multimessenger astronomy}
The detection and prompt localization of GW170817 event~\cite{PhysRevLett.119.161101} can be regarded as a monumental discovery that led to follow-up observations over the entire electromagnetic (EM) spectrum. This discovery resulted in the first extensive multi-messenger astronomical observing campaign~\cite{Abbott_2017} undertaken to follow up post-merger emissions from compact binary coalescence. The concurrent observations of such events via electromagnetic, neutrino, and gravitational wave (GW) detectors facilitate complementary measurements of phenomena that would otherwise remain inaccessible when observed independently. For example, the standard siren measurement of Hubble-Lema\^{i}tre constant $H_0$ from the GW data is independent of the cosmic distance-ladder method as opposed to that in astronomy with EM radiation. This provides a complementary measurement of $H_0$, which can be used to elucidate the Hubble-Lema\^{i}tre tension associated with the disparity between the measurements of Hubble constant~\cite{Riess_2019,2017, Abbott_2021} in the early and late universe.

\label{Different bands of EM emissions & related info}
Early observations of post-merger emissions of binary neutron star (BNS) sources 
can be used to constrain the physical models behind the internal mechanisms of these emission processes. 
For instance, there are different models proposed explaining the origin of the early blue emission of the kilonova associated with GW170817. Even though the model proposing radioactive decay of heavy elements in low-opacity ejecta being a theoretically motivated candidate~\cite{Arcavi_2018, Grossman_2014, Roberts_2011, Metzger_2010} fits the rise time curve well enough, there are also other models (for example, by cooling of shock-heated ejecta) which fit the decline of the emission equally well. In the case of GW1701817, the associated kilonova was detected ${\sim 11}$ hours after the merger, limiting the information about the rise time of the kilonova, particularly in the ultraviolet band~\cite{Arcavi_2018}. Capturing the rise time of the emission light curves by early detection may provide a useful measure to differentiate between the kilonova origin models. The gamma-ray burst (GRB) GRB 170817A was detected independently $\sim 1.7$ seconds after the trigger time of GW170817, with studies later confirming its association with the BNS merger~\cite{Abbott_2017_GRB}. This association confirmed BNS mergers as the progenitors of at least certain short GRBs~\cite{Abbott_2017_GRB}. 
The simultaneous detection of GWs and GRB may provide remarkable insights into the central engine of short gamma-ray bursts (SGRBs). The time delay between the GW and GRB events ($\sim 1.7$ secs as in the case of GW170817) offers valuable information into the underlying physics and intrinsic processes within the core, including the formation of a remnant object and the subsequent jet. It is expected that the EM waves and GW must have identical propagation speeds. The time delay between the GW and GRB events can be used to put constraints on the deviation of the speed of gravity waves from the speed of light, hence allowing for the tests of fundamental laws of physics~\cite{Abbott_2017_GRB}. Radio emissions enable tracing the fast-moving ejecta from BNS coalescence, providing insights into explosion energetics, ejecta geometry, and the merger environment~\cite{Geng_2018, Ghirlanda_2019}. Meanwhile, X-ray observations are crucial for determining the energy outflow geometry and system orientation relative to the observer's line of sight~\cite{Troja_2017}. The post-merger transients possess observational time scales ranging from a few seconds to weeks, owing to their potential to harness radiation across the entire EM spectrum. Detecting EM counterparts to $\sim 50$ BNS mergers may enable determining $H_0$ with $2\%$ fractional uncertainty~\cite{Chen_2018}, sufficient to verify the presence of any systematic errors in the local measurements of $H_0$. Even in the absence of potential EM counterparts, the accurate localization of compact binary coalescence (CBC) sources allows for dark siren measurements of $H_0$~\cite{1986Natur.323..310S, Finke_2021, Soares_Santos_2019}. For instance, observations from more than $\sim 50$ BNS dark sirens may be required to obtain $H_0$ measurements with $6\%$ fractional error~\cite{Chen_2018}. Thus, pinpointing these mergers is key in providing promising probes of fundamental physics, astrophysics, and cosmology. 

\label{Rapid PE for EM follow-up & Intro to LIGO India}
However, in order to accurately locate the EM counterparts and supplement regular follow-ups, more accurate and rapid 3D-localization of the sources using GW observations is of utmost importance. With imminent improvements in the detector sensitivities for future observing runs, there is an inevitable need for rapid Parameter Estimation (PE) methods. This arises from the fact that an increased bandwidth of detectors towards the lower cut-off frequencies would result in a momentous increase in the computational cost of Bayesian PE, hence affecting the prompt localization of the GW source. 
Currently, LIGO-Virgo-KAGRA Collaboration (LVK)~\cite{Abbott_2020_Prospects} uses a Bayesian, non-MCMC-based rapid sky localization tool, known as {\texttt{BAYESTAR}}~\cite{Singer_2016}, to locate (posterior distributions over sky location parameters, $\alpha$, and $\delta$) CBC sources within tens of seconds following the detection of the corresponding GW signal. However, as shown by Finstad \textit{et al.}~\cite{Finstad_2020}, a full Bayesian analysis including both intrinsic and extrinsic parameters can significantly increase the accuracy of sky-localization (by ${\sim {14 \,\text{deg}^2}}$ in their analysis) of the CBC sources. This is of primary significance in reducing the telescope survey area \& time for locating the EM counterparts, making rapid Bayesian PE-based methods a preferable as well an evident necessity. Nevertheless, since \texttt{BAYESTAR} can construct skymaps in the order of a few tens of seconds, it would be an interesting case to test if the \texttt{BAYESTAR} skymap samples can be used as priors for sky location estimates in rapid Bayesian PE methods. This scheme might enable a more accurate localization measurement of the source. Future improvements in detector noise sensitivities shall lead to an increase in the detector ``sensemon range''\footnote{\textit{``sensemon range"} is defined as the radius of a sphere of volume in which a GW detector could detect a source at a fixed SNR threshold, averaged over all sky locations and orientations~\cite{Virgo_status_research_gate}. For lower redshifts ($z \lesssim 1$), the \textit{sensemon range} is approximately equal to the horizon distance divided by $2.264$.~\cite{Chen_2021}}~\cite{lsc2010sensitivity, Chen_2021, PhysRevD.47.2198}. 
Hence, higher BNS detections are expected from distances further away than current ranges~\cite{Pankow_2020}. It is projected that $\sim 180$ BNS events could be detected in future O5 run~\cite{kiendrebeogo2023updated}. Hence, it is judicious to start the EM follow-up right from the merger epoch, which would require early warning alerts~\cite{Sachdev_2020, Magee_2021} for the EM telescopes in future runs. 
Given the finite observational resources of EM facilities allotted for the GW localized regions, it is important to prioritize the follow-up based on the chirp-mass estimates of GW events, as suggested by Margalit \textit{et al.}~\cite{Margalit_2019}. This can be facilitated by a rapid Bayesian PE analysis, which, in addition to the sky localization, more importantly, provides a significantly accurate estimation of chirp mass for these compact binary systems. However, there are studies~\cite{Biscoveanu:2019ugx} which show that the chirp mass can be estimated accurately (with uncertainty no larger than $\sim 10^{-3}\, M_{\odot}$)  in low-latency searches. Although the mass ratio and effective aligned spin estimates may be severely biased in this case. With the addition of new detectors in the ground-based detector network, it becomes important to observe improvements expected in the sky localization and source parameter estimations of these compact binary sources.

LIGO-Aundha~\cite{LIGO_India, LIGO_India_location, UNNIKRISHNAN_2013, unnikrishnan2023ligoindia} (hereafter, `A1') is set to join the network of ground-based GW detectors in the early years of the next decade. Based on our experience with currently operational interferometric GW detectors, we expect A1 to go through multiple commissioning rounds, resulting in incremental improvement to its noise sensitivity, effective bandwidth, and duty cycle. We expect that it will eventually attain the target \texttt{aLIGO} A+ design sensitivity~\cite{Abbott_2020_Prospects, Noise_curves_used_for_Simulations_in_the_update_of_the_Observing_Scenarios_Paper} (sometimes referred to as the `O5' sensitivity in this paper) and acquire higher up-times, resulting in duty cycles that are commensurate with the stable operations of other detectors in the IGWN.
To model this progression, we consider A1 to have two distinct noise sensitivities: one at conditions that mirror the fourth science run (O4) of the LIGO Hanford and Livingston detectors, simulating an early commissioning stage, and the other at the A+ design sensitivity of the detector. Additionally, we consider various duty cycles of the A1 detector at the sensitivities mentioned above for a comprehensive analysis. 

By the time A1 begins its maiden science run, the current detectors namely LIGO-Hanford (H1) and LIGO-Livingston (L1)~\cite{Advanced_LIGO_2015}, are likely to reach their A+ design sensitivities or beyond; meanwhile Virgo (V1)~\cite{Acernese_2014} and KAGRA (K1)~\cite{akutsu2020overview} are also expected to reach their target design sensitivities. LIGO-Aundha (A1) would add significantly longer baselines to the existing GW network and also contribute to increasing the network SNR~\cite{Schutz_2011, Saleem_2021}, leading to better sky localizations of GW sources along with improvements in the estimations of source parameters. A number of studies done previously have addressed the localization capabilities of GW networks which include analytical studies~\cite{Fairhurst_2011, Wen_2010} as well as simulations using \texttt{BAYESTAR}, \texttt{LalInference} or other localization algorithms ~\cite{Pankow_2018, Chen_2017, Pankow_2020, Singer_2014, Rodriguez_2014}. 

\begin{figure}[htbp]
    \centering
    \hspace{-8mm}
    \includegraphics[width=0.5\textwidth, height=0.47\textwidth, clip=True]{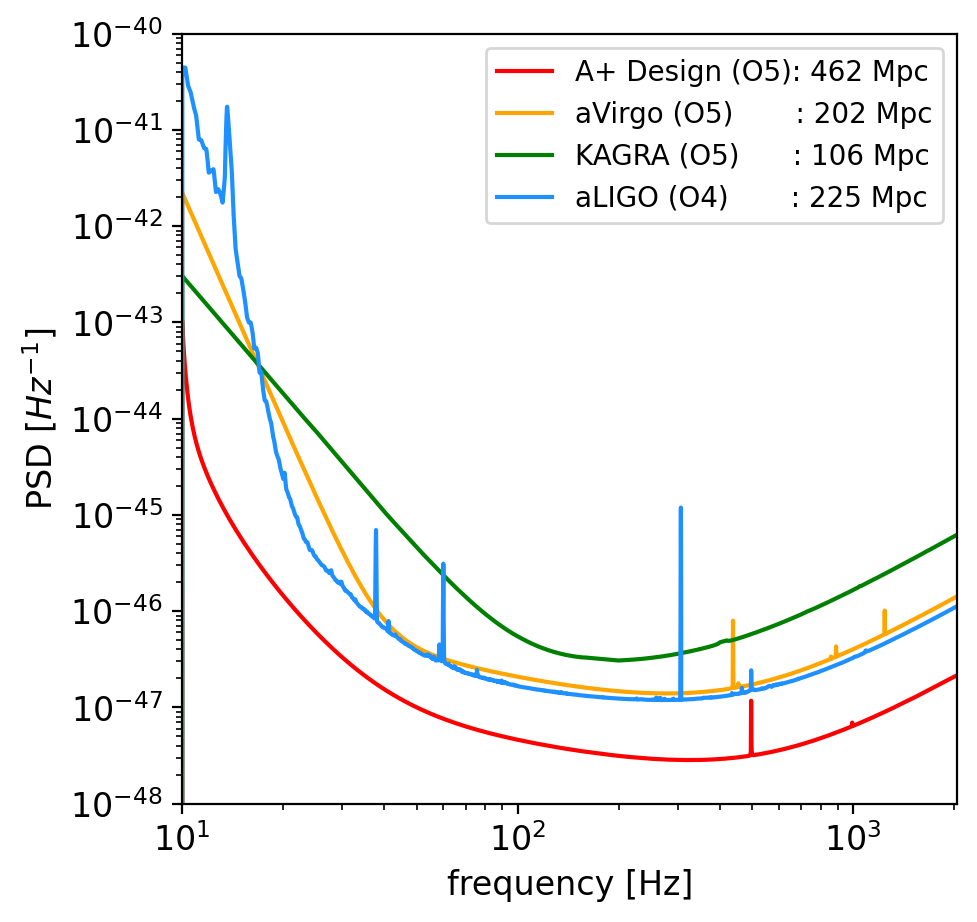}
    \caption{The noise Power Spectral Density (PSD)  for GW detectors. L1 and H1 are at \texttt{aLIGO A+ Design Sensitivity}(O5). V1 and K1 are at at \texttt{aVirgo}(O5) and \texttt{KAGRA(80 Mpc)} (O5) sensitivities respectively. We study A1 at two different sensitivities: first at \texttt{aLIGO} (O4) and then at \texttt{aLIGO A+ Design Sensitivity} (O5), respectively. The distances quoted here are the ``sensemon'' ranges for the respective detectors for a $1.4M_\odot+1.4M_\odot$ BNS at $\rho_{\text{th}}=6$. Note that the sensemon ranges for these PSD estimates presented in \cite{Abbott_2020_Prospects}, \cite{Noise_curves_used_for_Simulations_in_the_update_of_the_Observing_Scenarios_Paper} correspond to a detection signal-to-noise ratio threshold of 8 in a single detector.}
    \label{fig:PSD}
\end{figure}

\label{Our study & related info}
In this article, we aim to provide an illustration of the contribution A1 shall make to the current network of terrestrial GW detectors with a focus on improvements in sky localization capabilities for BNS mergers. 
We focus on the scenarios where the localization uncertainties are of the orders that enable a potential EM follow-up. This is possible when the source is localized by three or more detectors. 
We perform a full Bayesian Parameter Estimation using a rapid PE method developed by Pathak \textit{et al.}~\cite{pathak2022rapid, pathak2023prompt}. The method enables us to perform rapid Bayesian PE for BNS systems from a lower seismic cut-off frequency $f_{\text{low}} = 10 \hspace{0.5mm}\Hz$, hence allowing for the bandwidths (especially at lower frequencies) that would be typical of the future (\text{O5} \& beyond) observing runs. Analyzing CBCs from ${f_{\text{low}} = 10 \hspace{0.5mm}\Hz}$ increases the number of cycles in the frequency band and leads to an improvement in signal-to-noise ratio (SNR) accumulation and information content of the event. 

The paper is organized as follows: We describe the simulation study in Section~\ref{Simulation Section}. The section describes the detector network, duty cycles, simulated injection sources, and the Bayesian Inference method adopted here. Section~\ref{analysis of simulated events} outlines the priors and meshfree parameter configurations employed in the analysis. We summarise our localization results in Section~\ref{sky localization results} and Section~\ref{Subthreshold Events in LIGO-Aundha}. In Section~\ref{GWTC-Events}, we present a case of sky localization areas achieved for events from the Gravitational Wave Transient Catalog (GWTC) in the presence of A1, taking real noise into account. We also explore the effect of an additional detector in resolving the degeneracy between luminosity distance and inclination angle parameters. Finally, we conclude in Section~\ref{Conclusion_section} and describe possible improvements and discussion regarding this study in Section~\ref{Discussion}. 

\section{Simulations}
\label{Simulation Section}

The study of the localization capabilities of a GW detector network can be broadly characterized by the intrinsic source parameters (e.g. component masses)~\cite{Pankow_2018, Pankow_2020} of binary systems, detector sensitivities, and duty cycles of individual detectors~\cite{Schutz_2011,Pankow_2020}. We hereby discuss these aspects in relation to our work. 

\subsection{Detector Networks}
\label{describing the Heterogeneous network}
The ground-based GW detector network is comprised of detectors with different sensitivities. The heterogeneous nature of the noise Power Spectral Density (PSD) curves of detectors plays an important role in deciding the localization abilities of the network. Hence, we assume the ground-based GW detectors to be at different noise sensitivities for our analysis. The first two LIGO detectors L1 and H1 are configured to \texttt{aLIGO A+ Design Sensitivity}
PSDs~\cite{Advanced_LIGO_2015, Abbott_2020_Prospects}. The detector V1 is taken to be at its projected \texttt{aVirgo} PSD~\cite{Acernese_2014}. Meanwhile, the K1 detector is assumed to be at \texttt{KAGRA (80 Mpc)} O5 design sensitivity~\cite{akutsu2020overview, PhysRevD.88.043007, Somiya_2012}. 

To study the improvements in localization capabilities of the network with the addition of A1, we analyze the A1 detector at two different sensitivities. We first consider the case where A1 is in its initial operating phase at \texttt{aLIGO} O4 sensitivity. In the second case, we take A1 to be at \texttt{aLIGO A+ Design Sensitivity} configuration, marking the target sensitivity it shall achieve after undergoing staged commissioning over time. All the above-mentioned PSDs can be found in~\cite{Noise_curves_used_for_Simulations_in_the_update_of_the_Observing_Scenarios_Paper}. The aforementioned noise sensitivity curves are shown in Fig.~\ref{fig:PSD}. These sensitivity configurations, in conjunction with the duty cycles taken into account, shall provide a more comprehensive approach in presenting a science case for the addition of A1 to the current GW detector network. 

\subsection{Duty Cycles}
\label{describing the duty cycles}
The duty cycle of a detector/network is defined as the fraction of time for which it successfully collects data of scientific significance during an observing run~\cite{Saleem_2021}. The duty cycle for a detector depends on the specific phase in which the detector is relative to its intended target operating configuration. Additionally, environmental effects also play a role in affecting the detector duty cycle during an observation run. No detector can practically acquire science-quality data at all times. This translates to detectors working at different duty cycles depending on the commissioning of the detectors.

To understand the impact of duty cycles on the localization of BNS sources by a network, we {\it{assume}} the following three cases as suggested by Pankow \textit{et al.}~\cite{Pankow_2020} representing different stages of a detector's operation:

\begin{enumerate}
    \item [1)] 20\% duty cycle: A representative of the early stages of commissioning and engineering runs, resulting in reduced operational time.
    
    \item [2)] 50\% duty cycle: A representative of unresolved technical issues with the instrumental setup as well as signaling challenges like suboptimal environmental conditions.
    
    \item[3)] 80\% duty cycle: A representative of a detector operating near the possible target operating point. 
\end{enumerate}

Consider a GW network with $N$ detectors. Over the course of an observing run, there can be a $k$ number of detectors ($N_{\text{min}} \leq k\leq N$) participating in data collection, depending on their duty cycles. Here, $N_{\text{min}}$ represents the minimum number of detectors assumed to be participating in the observation of a GW event. These $k$ participating detectors can comprise different subnetworks of distinct detectors. For instance, a GW network with $N=5$ detectors may have only $k=4$ detectors in operation. Subject to which detector is out of operation, there can be $\binom{N}{k}=5$ different subnetworks of distinct detectors. Depending on the duty cycles of individual detectors, the effective duty cycle of a subnetwork can be evaluated. Out of a total of $N$ detectors in the network, we assume a set $\boldsymbol{m}$ composed of all the detectors participating in data collection and a set $\boldsymbol{n}$ comprising detectors that are out of operation (possibly due to maintenance) during this observation period. We represent the probability ($p$) of being in operation defined by a given duty cycle (i.e. $p=0.5$ for $50\%$ duty cycle). The probability representing the effective duty cycle of a subnetwork ($p_{\text{eff}}$) is given as
\begin{equation}
\label{equation_duty_cycle}
    p_{\text{eff}} = \prod_{m_i} \ \prod_{n_j} p_{m_{i}} (1 - p_{n_{j}}),
\end{equation}
where $p_{m_{i}}$ and $p_{n_{j}}$ are the duty cycles of the $i$th detector in set $\boldsymbol{m}$ and $j$th detector in set $\boldsymbol{n}$ respectively. For instance, consider a network of $N=4$ detectors, namely L1, H1, V1, and K1. The probability of being in operation for individual detectors is given by $p_{\text{L1}}$, $p_{\text{H1}}$, $p_{\text{V1}}$, and $p_{\text{K1}}$ representing their respective duty cycles. There may be a scenario where any one of these four detectors may get out of operation due to maintenance or environmental causes. Depending on which detector is not in observation mode, there can be $\binom{4}{3}=4$ subnetworks, comprising of $k=3$ detectors participating in data acquisition. 
To evaluate the probability of being in operation $p_{\text{eff}}$ for one of the subnetworks consisting of say H1, V1, and K1 detectors (this is the case when \texttt{L1} is not in operation) we have
\begin{equation*}
    p_{\text{eff}}\Big|_{\text{H1V1K1}} = p_{\text{H1}} \cdot p_{\text{V1}} \cdot p_{\text{K1}} \cdot (1 - p_{\text{L1}})
\end{equation*}
where, H1, V1, K1 $\in \textbf{m}$ and L1 $\in \textbf{n}$ respectively. The effective duty cycle for $N$-detector network is evaluated as the probability obtained by adding the probabilities of being in operation for all subnetworks over $k \leq N$ number of participating detectors. 

In this study, we consider the L1, H1, V1, and K1 detectors operating near their target operating point at 80\% duty cycle each and are fixed during the analysis. As A1 is expected to join this network by the early 2030s, we aim to show how the addition of A1 improves the localization capabilities of the GW network. We vary duty cycles for the A1 detector, simulating the cases for various phases of its configuration relative to its target operating point. 
Single detectors are nearly omnidirectional due to the structure of the antenna pattern functions - leading to poor source localization. For a two-detector network, solving for the direction in the sky corresponding to a fixed time-delay between the coalescence times recorded at the detectors leads to a `ring' like pattern on the sky, and as such, events localized with two detectors are generally not useful enough for EM follow-ups. Hence, we do not include the cases with $k \leq 2 $ in our analysis. For the purpose of this study, we assume $3 \leq k \leq N$ (i.e. $N_{\text{min}} = 3$) and evaluate the network duty cycles accordingly.  
The impact of A1 as an addition to the second-generation GW network consisting of L1, H1, V1, and K1 detectors is presented by the implementation of varying duty cycles (20\%, 50\%, 80\%) in conjunction with different detector sensitivities (\texttt{aLIGO} O4 and \texttt{aLIGO A+ Design Sensitivity}) for A1 detector. 

\subsection{Injected BNS sources}
\label{motivating for SNR range choice & injection parameters}

The remarkably high signal-to-noise ratio (SNR) due to the fortunate proximity of the GW170817 event enabled its effective localization and multi-messenger efficacies. Since the number of such `golden events' is expected to be low even in future observing runs~\cite{Schutz_2011, Baibhav_2019}, it becomes increasingly important to study a network's ability to localize such events. 
Hence, we aim to focus on studying the impact of the addition of the LIGO-Aundha detector in localizing moderately high SNR events, which may lead to potential multi-messenger observations.  Taking this into consideration, we choose to generate 500 BNS events having an optimal network SNR in the range of $20$ to $25$ in the GW network comprising L1, H1, V1, and K1 detectors. We use these events for the purpose of our investigation. We would like to highlight that this is not a population study but focuses on possible improvements to the localization capabilities of a GW network with the addition of the LIGO-Aundha detector in accurately locating such `golden events'. An event is considered to be detected if the individual detector optimal SNR is greater than a threshold value of 6 $(\rho_{\text{th}}> 6)$ in at least two detectors. From all the generated BNS sources, an event must follow the detection criteria to be considered detected by a subnetwork/network. The effectiveness of a network with an additional A1 detector is studied against the four-detector network with L1, H1, V1, and K1  detectors. 

\label{mass and spin choice}
The intrinsic parameters, like component masses, spins, etc., affect the localization of CBC sources. The effective bandwidth, as defined in~\cite{Fairhurst_2011}, measures the frequency content of the signal. Effective bandwidth is one of the important factors affecting the localization of CBC sources~\cite{Pankow_2018}. The signals from BNS sources mostly span through the entire bandwidth of the ground-based detectors owing to their relatively small component mass values in comparison to binary black holes or neutron star-black hole sources. The mass ranges for the BNS are also narrow, leading to small variations in effective bandwidths. In fact, it has also been shown by Pankow \textit{et al.}~\cite{Pankow_2020} that the sky localization uncertainties for BNS systems are effectively independent of the population model of their component masses and spins. Hence, in order to simplify our simulations, we work with a particular choice of component source masses and spins. The tidal parameters are also expected to have a negligible effect on the source localization of these systems~\cite{Pankow_2018} and, therefore, are not included as source parameters.

The source-frame intrinsic parameters are chosen to have the maximum a \textit{posteriori} (MAP) values of the posterior samples obtained from the LIGO PE analysis of GW170817~\cite{Romero_Shaw_2020} using \texttt{BILBY}~\cite{Ashton_2019} Python package. The source frame component masses are ${m^{\text{src}}_{1} = 1.387 \:M_{\odot}}$, ${m^{\text{src}}_{2} = 1.326\:M_{\odot}}$, while the dimensionless aligned spin component parameters are ${\chi_{1z} \approx 1.29 \times 10^{-4}}$,  ${\chi_{2z} \approx 3.54 \times 10^{-5}}$. We choose the inclination angle $\iota=\pi/6$ \texttt{radians}. We set the polarization angle arbitrarily to $\psi=0$ \texttt{radians}. The sources are distributed uniformly in sky directions. We distribute the sources in luminosity distances corresponding to the redshifts following a uniform in comoving volume distribution up to a redshift of $\sim 0.14$, which is greater than the detection range of a detector at \texttt{aLIGO A+ Design Sensitivity} (O5) for a BNS with component masses $m^{\text{src}}_1$ and $m^{\text{src}}_2$. In addition to the sources that are to be detected with certainty, this limit also allows for events that are barely near the threshold in some detectors. We simulate uncorrelated Gaussian noise in each detector characterized by their associated PSDs respectively. The injection sources are generated with the \texttt{IMRPhenomD}~\cite{khan2016frequency} waveform model, and the source parameters are recovered using the \texttt{TaylorF2}~\cite{PhysRevD.49.1707, PhysRevD.59.124016, Faye_2012, PhysRevLett.74.3515} waveform model for the Bayesian PE analysis. Since BNS mergers are mostly inspiral-dominated in the LIGO-Virgo-KAGRA (LVK) detector band, the use of the \texttt{TaylorF2} waveform model sufficiently extracts the required information from strain data.

\subsection{Bayesian inference}
\label{Bayesian PE}
In order to estimate the parameters of GW sources, we use a Bayesian framework, where for a given waveform model $\boldsymbol{h}$ and data $\boldsymbol{d}$ from the detectors, the posterior distribution of the source parameters $\vec \Lambda$ can be estimated via Bayes' theorem:
\begin{equation}
    p(\vec \Lambda \mid \boldsymbol{d}) = \frac{\mathcal{L}(\boldsymbol{d}\mid \vec \Lambda)\: p(\vec \Lambda)}{p(\boldsymbol{d})}
\end{equation}
where ${\mathcal{L}(\boldsymbol{d}\mid \vec \Lambda)}$ is the likelihood function, $p(\vec \Lambda)$ is the prior over the source parameters ${\param \equiv \{ \vec \lambda, \vec \theta, t_{c} \}}$, and $p(\boldsymbol{d})$ is called evidence, which describes the probability of data given the model. Here $\vec \lambda$ represents the intrinsic parameters, whereas $\vec \theta$ denotes the extrinsic parameters. In principle, ${p(\vec \Lambda \mid \boldsymbol{d})}$ can be estimated by placing a grid over the parameter space $\vec \Lambda$, which for a typical compact binary coalescence (CBC) source described by a ${\sim 15}$ dimensional parameter space, would become practically intractable. In the case of a BNS system, it increases to $17$ dimensional space due to the addition of two tidal deformability parameters. Instead, stochastic sampling methods such as Markov chain Monte Carlo (MCMC)~\cite{Foreman_Mackey_2013} and Nested Sampling~\cite{skilling2006nested} are employed to generate representative samples from the posterior distribution ${p(\vec \Lambda\mid \boldsymbol{d})}$. However, this process still requires evaluating the likelihood function, which involves a computationally expensive step of generating model (template) waveforms at the proposed points by the sampler and calculating the overlap between these waveforms and the data. This computational cost is notably significant, especially for low-mass systems such as binary neutron star (BNS) events with lower cutoff frequency decreased to $10$ Hz. The situation is exacerbated by the enhanced sensitivity of detectors, resulting in a large number of in-band waveform cycles. Furthermore, incorporating additional physical effects can further escalate the computational burden of waveform generation. These factors have significant implications for the feasibility of promptly following up on EM counterparts of corresponding BNS systems.
As mentioned earlier, a high number of BNS detections are expected in the O5 runs; it would be prudent to prioritize the EM follow-ups given the limited observational resources. This underscores the importance of the development of rapid PE methods, which can efficiently estimate both intrinsic and extrinsic parameters. Various rapid PE methods have been proposed in the recent past. They broadly come under two categories: (i) ``likelihood-based'' approaches such as Reduced order models~\cite{canizares2013gravitational, canizares2015accelerated, smith2016fast, Morisaki_2020, morisaki2023rapid}, Heterodyning (or Relative Binning)~\cite{Venumadhav2018, cornish2021heterodyned, Finstad_2020, islam2022factorized}, and other techniques such as RIFT~\cite{https://doi.org/10.48550/arxiv.1805.10457}, simple-pe~\cite{fairhurst2023fast}, multibanding~\cite{Morisaki:2021ngj} (ii) ``likelihood-free'' approaches which aim to directly learn the posteriors employing Machine-learning techniques such as deep learning, normalizing flows, and variational inference as well ~\cite{chua2019reduced, chua2020learning, Green_2020, green2020complete, gabbard2022bayesian}. 

In this work, we use a likelihood-based rapid PE method developed by Pathak \textit{et al.}~\cite{pathak2022rapid, pathak2023prompt}, which combines dimensionality reduction techniques and meshfree approximations to swiftly calculate the likelihood at the proposed query points by the sampler. This algorithm is interfaced with {\sc{dynesty}}~\cite{speagle2020dynesty, sergey_koposov_2023_7600689}, a Python implementation of the Nested sampling algorithm to quickly estimate the posteriors distribution over the source parameters. In the forthcoming sections, we will first define the likelihood function and subsequently provide a concise overview of how the meshfree method expeditiously computes the likelihood at the sampler's proposed query points.

\subsubsection{Likelihood function}
Given a stream of data $d^{(i)}$ from the $i^{\text{th}}$ detector and a template $\tilde{h}^{(i)}(\vec \Lambda)$, under an assumption of uncorrelated noise across the detectors, the coherent network log-likelihood is given by 
\begin{multline}
\label{eq:multigenlikelihood}
\ln \mathcal{L}(\vec\Lambda) =  \sum_{i=1}^{N_{\text{d}}} {\langle \boldsymbol{d}^{(i)} \mid \tilde{h}^{(i)}(\vec\Lambda)\rangle} \\ 
- \frac{1}{2} \sum_{i=1}^{N_{\text{d}}} \left [ \| \tilde{h}^{(i)}(\vec\Lambda)\|^2 + \| \boldsymbol{d}^{(i)} \|^2 \right ]
\end{multline} 
where $\tilde{h}^{(i)}(\vec \Lambda)$ represents the frequency domain Fourier Transform (FT) of the signal $h^{(i)}(\vec \Lambda)$ and $N_d$ is the number of detectors. Here, the inner product is defined as 
\begin{equation}
    \langle x \mid y\rangle = 4\,\text{Re}\int_{0}^{\infty} \, df\,\frac{{\tilde{x}}(f)^{*}\,\,\tilde{y}(f)}{S_h(f)}
    \label{eq:inner_prod_def}
\end{equation}
In this paper, we focus on the non-precessing GW signal model, which can be decomposed into factors dependent on only intrinsic and extrinsic parameters as follows:
\begin{equation}
\label{eq:detframwaveform}
\begin{split}
\tilde{h}^{(i)}(\vec\Lambda) \equiv \tilde{h}(\vec\Lambda, t^{(i)}) 
    &= \mathcal{A}^{(i)} \tilde{h}_{+}(\vec \lambda, t^{(i)}), \\
    &= \mathcal{A}^{(i)}\, \tilde{h}_{+}(\vec \lambda)\, e^{-j\,2\pi f t_c} \, e^{-j\,2\pi f \Delta t^{(i)}}
\end{split}
\end{equation}
where $\mathcal{A}^{(i)}$, the complex magnitude of the signal depends only on the extrinsic parameters $\vec\theta \in \vec \Lambda$ through the antenna pattern functions, luminosity distance $d_L$, and the inclination angle $\iota$, and can be expressed as the following:
\begin{multline}
\mathcal{A}^{(i)} = 
    \frac{1}{d_L} \left[ \frac{1+\cos^2 \iota}{2}  F^{(i)}_{+}(\alpha, \delta, \psi) \right.\\
            \left. - \: j \cos \iota \ F^{(i)}_{\times}(\alpha, \delta, \psi) \right],
\end{multline}
$\Delta t^{(i)}$ corresponds to the time-delay introduced due to the relative positioning of the $i^{\text{th}}$ detector in relation to the Earth's center~\cite{pankow2015novel}, the ${F^{(i)}_{+}(\alpha, \delta, \psi)}$ and ${F^{(i)}_{\times}(\alpha, \delta, \psi)}$ are respectively the `plus' and `cross' antenna pattern functions of the $i^{\text{th}}$ detector, which are functions of right-ascension $\alpha$, declination $\delta$, and polarization angle $\psi$. The antenna pattern functions describe the angular response of the detector to incoming GW signals~\cite{detectorTensor}. 

In our analysis, we opt for the log-likelihood function marginalized over the coalescence phase~\cite{thrane_2019}. With $\tilde{h}^{(i)}(\vec \Lambda)$ given by Eq.~\eqref{eq:detframwaveform}, the expression of the marginalized phase likelihood is given by 
\begin{multline}
\label{eq:multiphaselikelihood}
    \left.\ln \mathcal{L}(\vec\Lambda \mid \boldsymbol{d}^{(i)})\right|_{\phi_c} 
    = \ln I_{0}\left[\left|\sum_{i=1}^{N_{d}}{{\mathcal{A}}^{(i)}}^{*}\, \langle \boldsymbol{d}^{(i)} \mid \tilde{h}_{+} (\vec\lambda, t^{(i)}) \rangle \right|\right] \\
    - \frac{1}{2}\sum_{i=1}^{N_{d}}\left[ \left|\mathcal{A}^{(i)}\right|^2 \sigma^2(\vec \lambda)^{(i)} + \| \boldsymbol{d}^{(i)} \|^2 \right];
\end{multline}
where $I_0(\cdot)$ is the modified Bessel function of the first kind and ${\vec z^{(i)}(\vec \lambda^n) \equiv \langle \boldsymbol{d}^{(i)} \mid \tilde{h}_{+}(\vec \lambda, t^{(i)}) \rangle}$ is the complex overlap integral, while ${\sigma^2(\vec\lambda)^{(i)} \equiv \langle \tilde{h}_{+}(\vec \lambda, t^{(i)}) \mid \tilde{h}_{+}(\vec \lambda, t^{(i)}) \rangle}$ is the squared norm of the template $\: \tilde{h}_{+} (\vec\lambda)$. ${\sigma^2(\vec\lambda)^{(i)}}$ depends on the noise power spectral density (PSD) of the $i^{\text{th}}$ detector. 
The squared norm of the data vector, $\| \boldsymbol{d}^{(i)} \|^2$, remains constant throughout the PE analysis and, hence, does not affect the overall `shape' of the likelihood. Consequently, it can be excluded in the subsequent analysis. Note that the marginalized phase likelihood will not be an appropriate choice for systems with high precession and systems containing significant power in subdominant modes~\cite{Pratten_2021}.

\subsubsection{Meshfree likelihood interpolation}
The meshfree likelihood interpolation, as outlined in~\cite{pathak2022rapid, pathak2023prompt}, comprises two stages: (i) Start-up stage, where we generate radial basis functions (RBF) interpolants of the relevant quantities and (ii) Online-stage, where the likelihood is calculated by evaluating the interpolants at the query points proposed by the sampler. Let's briefly discuss both stages.

\begin{itemize}
    \item \label{start-up stage}\textbf{Start-up stage}: First, we generate $N$ RBF interpolation nodes in the intrinsic parameter space ($\mathcal{M}$, $q$, $\chi_{1z}$, and $\chi_{2z}$ in this context). The center $\vec \lambda^{\text{cent}}$ around which these interpolation nodes are positioned is determined by optimizing the network-matched filter SNR, starting from the best-matched template or trigger $\vec \lambda^{\text{trig}}$ and $t_{\text{trig}}$ identified by the upstream search pipelines~\cite{GstLAL_2010, cannon2021gstlal, usman2016pycbc}. For simulated systems, the injection parameter is taken as the central point for node placement. We employ a combination of Gaussian and uniform nodes, where the Gaussian nodes are sampled from a multivariate Gaussian distribution (MVN) with a mean of $\vec \lambda^{\text{cent}}$ and a covariance matrix calculated using the inverse of the Fisher matrix evaluated at $\vec \lambda^{\text{cent}}$. A hybrid node placement approach ensures that nodes are positioned near the peak of the posterior, where higher accuracy in likelihood reconstruction is necessary. Once the nodes $\vec \lambda^{n}$ are generated, we efficiently compute the time-series ${\vec z^{(i)}(\vec \lambda^{n}) \equiv z^{(i)}(\vec \lambda^{n}, t_c)}$ using the Fast Fourier Transform (FFT) circular correlations, with $t_c$ being uniformly spaced discrete-time shifts within a specified range ($ \pm 150$ ms\footnote{This range should be larger than the maximum light travel time between two detectors.}) around a reference coalescence time $t_{\text{trig}}$. During this calculation, we set $\Delta t^{(i)} = 0$ for overlap time series, handling extra time offsets introduced due to sampling in the sky location parameters during the online stage.
    Similarly, we compute the template norm square $\sigma^2(\vec \lambda^n)^{(i)}$ at the RBF nodes $\vec \lambda^n$. We then stack the time series (row-wise) and perform Singular Value Decomposition (SVD) of the resulting matrix, producing a set of basis vectors spanning the space of $\vec z^{(i)}(\vec \lambda^n)$:
    \begin{equation}
    \label{eq:svdbasistimeseries}
    \vec z^{(i)}(\vec \lambda^{n}) = \sum_{\mu = 1}^N\, C^{n (i)}_{\mu}\ \vec u^{(i)}_{\mu}
    \end{equation}
    where the SVD coefficients $C^{n (i)}_{\mu}$, smooth functions of $\vec \lambda^{n}$ within the sufficiently narrow boundaries encompassing the posterior support, can be interpolated over the $\vec \lambda$ using a linear combination of RBFs and monomials~\cite{doi:10.1142/6437}: 
    \begin{equation}
        \label{eq:rbfcoeff}
        C^{q (i)}_{\mu} = \sum_{n=1}^N\, a^{(i)}_{n}\, \phi(\|\vec \lambda^q - \vec \lambda^{n}\|_2) + \sum_{j = 1}^{M}\, b^{(i)}_{j}\, p_j(\vec \lambda^q)
    \end{equation}
    where $\phi$ is the RBF kernel centered at ${\vec \lambda^{n} \in \mathcal{R}^d}$, and $\left \{  p_{j} \right \}$ denotes the monomials that span the space of polynomials with a predetermined degree $\nu$ in $d$-dimensions. Since the coefficients are only known at $N$ RBF nodes $\vec \lambda^n$, we impose $M$ additional conditions of the form ${\sum_{j=1}^M a^{(i)}_j p_j(\vec \lambda^q) = 0}$ to uniquely solve for the coefficients $a_n$ and $b_j$ in the Eq.~\eqref{eq:rbfcoeff}. Furthermore, it turns out that only ``top-few'' basis vectors are sufficient to reconstruct $\vec z^{(i)}(\vec \lambda^q)$ at minimal reconstruction error. Consequently, we generate only top-$\ell$ meshfree interpolants of $C_{\mu}^{q(i)}$ where $\mu = 1,....,\ell$, where $\ell$ can be chosen based on the singular value profile. Similarly, we express $\sigma^2(\vec \lambda^q)$ in terms of RBFs and monomials, treating them as smoothly varying functions over the interpolation domain. Finally, we have uniquely constructed the $\ell + 1$ RBF interpolants, which are to be used in the online stage. 

    \item \textbf{Online stage}: In the online stage, we rapidly compute interpolated values of $C^{q (i)}_{\mu}$ and $\sigma^2(\vec \lambda^q)^{(i)}$ at any query point $\vec \lambda^q$ within the interpolation domain. Subsequently, we determine the corresponding $\vec z^{(i)}(\vec \lambda^q)$ using Eq.\eqref{eq:svdbasistimeseries}. Rather than generating the entire time series, we focus on creating $\vec z^{(i)}(\vec \lambda^q)$ with around $\mathcal{O}(10)$ time samples centered around the query time $t^{q (i)}$, which contain the additional time-offset $\Delta t^{(i)}$. We fit these samples with a cubic spline, from which we calculate $\vec z^{(i)}(\vec \lambda^q)$ at the query time $t^{q (i)}$. Similarly, we compute the interpolated value of $\sigma^2(\vec \lambda^q)^{(i)}$. Finally, we integrate these interpolated values with the factors related to extrinsic parameters, as outlined in Eq.\eqref{eq:multiphaselikelihood}, to compute the interpolated likelihood $\ln \mathcal{L}_{\text{RBF}}$.
\end{itemize}

\section{Analysis of Simulated Events}
\label{analysis of simulated events}
As discussed previously in Section~\ref{motivating for SNR range choice & injection parameters}, we create injections with fixed source-frame masses and dimensionless aligned spin component parameters. However, the detector-frame parameters (masses) for these events vary according to their associated redshifts. We define the intrinsic detector-frame parameters by ${\vec\lambda = (\mathcal{M}_{\text{det}}, q, \chi_{1z}, \chi_{2z})}$. Similarly, the injected intrinsic parameters in the detector frame are denoted as $\vec\lambda^{\text{cent}}$. To perform Bayesian PE for each event, we first generate $N_{\text{nodes}} = 800$ RBF nodes as described in Section~\ref{start-up stage}. We sample $20\%$ of the total RBF nodes ($N_{\text{Gauss}} = 160$) from a multivariate Gaussian distribution $\mathcal{N}(\vec \lambda^{\text{cent}}, \mathbf{\Sigma})$, where $\vec\lambda^{\text{cent}}$ is the mean and $\mathbf{\Sigma}$ is the covariance matrix obtained from inverse of the Fisher matrix $\mathbf{\Gamma}$ around the center $\vec\lambda^{\text{cent}}$ using the \texttt{gwfast}~\cite{Iacovelli_2022} python package. The remaining $80\%$ of the total RBF nodes ($N_{\text{Unif}} = 640$) are sampled uniformly from the ranges provided in Table \ref{tab:priordistr}. We choose $\phi = \exp(-(\epsilon r)^2)$ as the Gaussian RBF kernel in our analysis, with $\epsilon$ being the shape parameter. For the purpose of this analysis, we use $\epsilon=10$, monomial terms with degree $\nu=7$ and $l=20$ top basis vectors for reconstructing the time-series in Eq.~(\ref{eq:svdbasistimeseries}). After the successful generation of interpolants, the likelihood function can be evaluated using $\ln \mathcal{L}_{\text{RBF}}$ by sampling the ten-dimensional parameter space $\vec \lambda$  using the \texttt{dynesty} sampler. The sampler configuration is outlined as follows: \texttt{nlive} $=500$, \texttt{walks} $=100$, \texttt{sample} = ``rwalk'', and \texttt{dlogz} $=0.1$. These parameters play a critical role in determining both the accuracy and the time required for the nested sampling algorithm to converge. In this context, the parameter \texttt{nlive} represents the number of live points. Opting for a larger value of \texttt{nlive} leads to a more finely sampled posterior distribution (and consequently, the evidence), but it comes at the cost of requiring more iterations to achieve convergence. The parameter \texttt{walks} specifies the minimum number of points necessary before proposing a new live point, \texttt{sample} indicates the chosen approach for generating samples, and \texttt{dlogz} represents the proportion of the remaining prior volume's contribution to the total evidence. In this analysis, \texttt{dlogz} $=0.1$ serves as a stopping criterion for terminating the sampling process. For a more comprehensive understanding of dynesty's nested sampling algorithm and its practical implementation, one can refer to the following references ~\cite{speagle2020dynesty, sergey_koposov_2023_7600689}.

The prior distribution for $\vec \lambda$, along with the associated parameter space boundaries, are presented in Table \ref{tab:priordistr}. The prior distributions for the extrinsic parameters ($\alpha, \delta, V_{\text{com}}, \iota, \psi, t_c$) and their respective parameter space boundaries are also presented in Table~\ref{tab:priordistr}. To evaluate the Bayesian posteriors of source parameters, we sample over the entire ten-dimensional parameter space involving four intrinsic and six extrinsic source parameters. This ensures accounting for the correlations between parameters. However, the focus of this study lies in discussing the sky localization uncertainties obtained from the posteriors over $\alpha$ and $\delta$ parameters.

\begin{table}[hbt]
\def\arraystretch{1.5}
\begin{ruledtabular}
\begin{tabular}{lcl}
 Parameters     &   Range    &   Prior distribution \\
\hline
 $\mathcal{M_{\text{det}}}$  &   $[\mathcal{M}^{\text{cent}}_{\text{det}} \pm 0.0001]$ &   $\propto \mathcal{M_{\text{det}}}$ \\
 $q$            &   $[q^{\text{cent}} \pm 0.07]$ & $ \propto \left [ (1 + q)/q^3 \right ]^{2/5}$     \\
 $\chi_{1z, 2z}$   &   $[\chi_{1z}^{\text{cent}} \pm 0.0025]$   & Uniform  \\
 $V_{\text{com}}$          & $[\sim16.9e3, \sim8.9e8]$                & Uniform \\
 $t_c$          & $t_{\text{trig}} \pm 0.12$& Uniform\\
 $\alpha$       & $[0, 2\pi]$               & Uniform\\
 $\delta$       & $\pm \pi/2$       & $\sin^{-1} \left [ {\text{Uniform}}[-1,1]\right ]$\\
 $\iota$        & $[0, \pi]$                & Uniform in $\cos \iota$\\
 $\psi$         & $[0, 2\pi]$               & Uniform angle\\
\end{tabular}
\end{ruledtabular}
\caption{Prior parameter space over the ten-dimensional parameter space $\vec \Lambda$.\footnote{For the case where $q^{\text{cent}} - 0.07 < 1$, we sample $q$ from the range $[1, 1.14]$ respecting the same prior width.}}
\label{tab:priordistr}
\end{table}

In accordance with the previous discussion in Section~\ref{Bayesian PE}, we perform PE for the simulated events with different subnetworks of a GW network to take into account the effect of duty cycles. For instance, in the case of a network with L1, H1, V1, K1, and A1, there can be $10$ different subnetworks consisting of three distinct detectors ($k=3$), and $5$ different subnetworks of four distinct detectors ($k=4$) taking observations depending on the duty cycles. In addition to these,  there is a subnetwork consisting of all the five detectors for $k=5$ case.

Bayesian PE analyses are performed for the events detected in each of these subnetworks. The total number of subnetworks for all $3\leq k \leq 5$ is 16 for the five detector networks comprising of L1, H1, V1, K1, and A1 detectors. The exercise is repeated for two cases: 
\begin{enumerate}
    \item [(i)] Keeping A1 at \texttt{aLIGO} O4 noise sensitivity in the GW network. Here, the A1 sensitivity is close to \texttt{aVirgo} (O5) sensitivity (Refer Fig.~\ref{fig:PSD}).

    \item [(ii)] Setting A1  at \texttt{aLIGO A+ Design Sensitivity} (O5) in the GW network. In this case, the A1 detector would be at the same sensitivity as the other two LIGO detectors.
\end{enumerate}

We represent the network with $N=5$ detectors as the L1H1V1K1A1 network and, similarly, the network with $N=4$ detectors as the L1H1V1K1 network. Using the \texttt{ligo-skymap}~\cite{Singer_2016} utility, we compute the $90\%$ credible sky localization areas $\Delta \Omega_{90\%}$ (in sq. deg) from the posterior samples over right ascension ($\alpha$) and declination ($\delta$) obtained from Bayesian PE.

\begin{figure*}[htbp]
    \centering
    \begin{subfigure}{0.49\linewidth}
        \centering
        \includegraphics[height=2.3in, width=\linewidth]{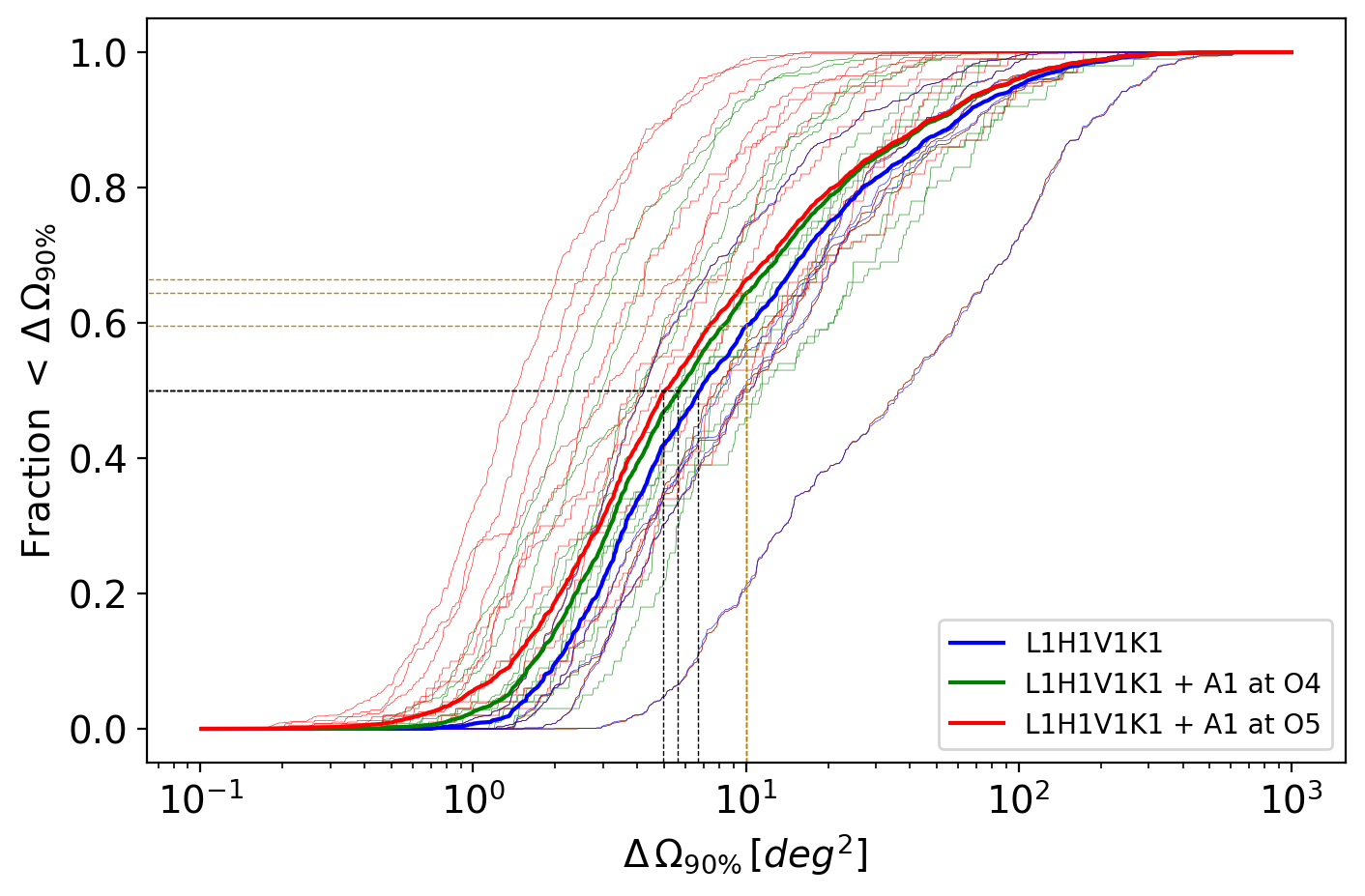} 
        \caption{A1 at 20\% duty cycle}
        \label{fig: duty cycle plot1}
    \end{subfigure}%
    \hfill
    \begin{subfigure}{0.49\linewidth}
        \centering
        \includegraphics[height=2.3in, width=\linewidth]{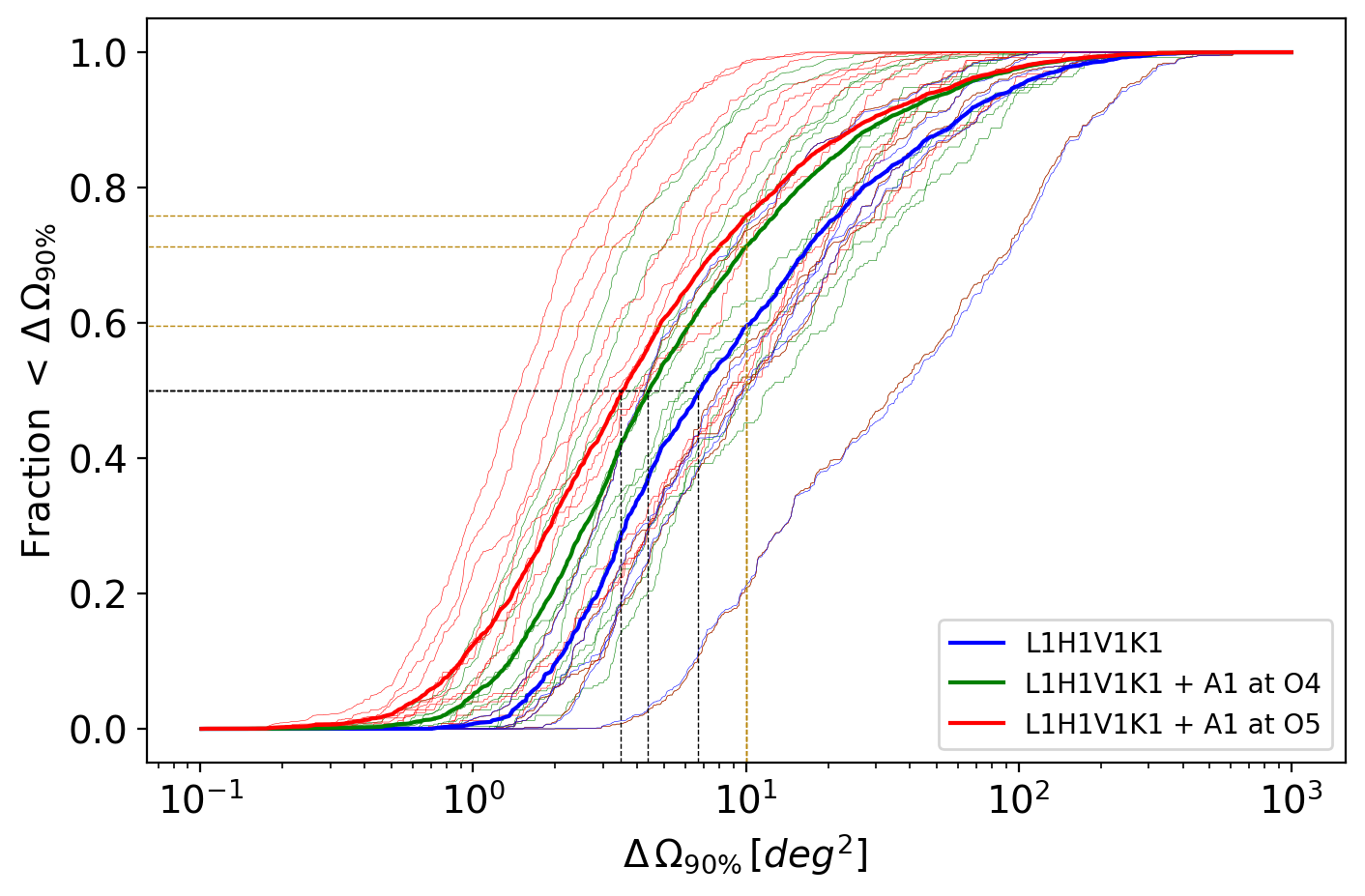} 
        \caption{A1 at 50\% duty cycle}
        \label{fig:duty cycle plot2}
    \end{subfigure}
    \hfill
    \\
    \begin{subfigure}{0.49\linewidth}
    \vspace{5mm}
        \centering
        \includegraphics[height=2.3in, width=\linewidth]{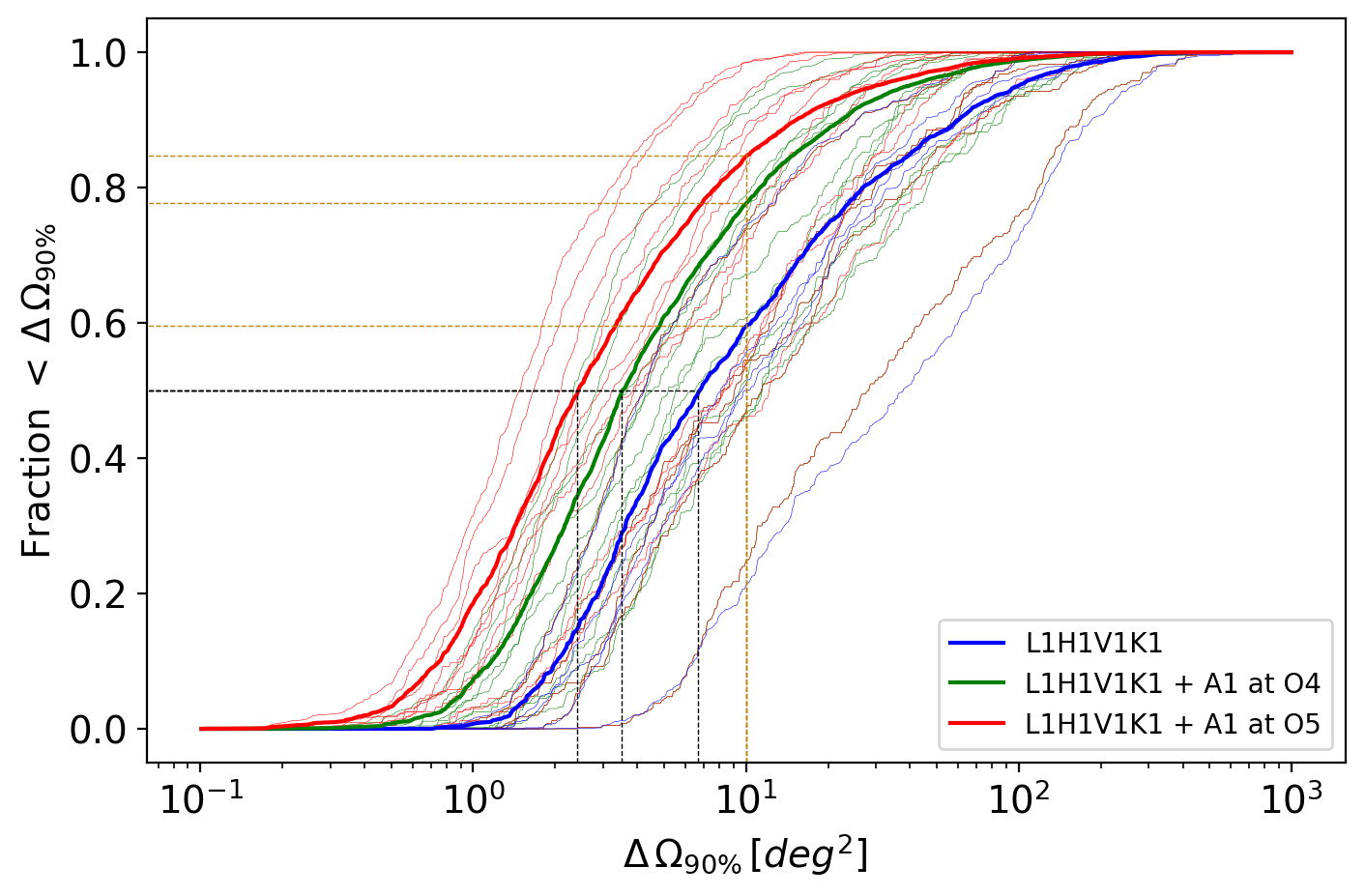} 
        \caption{A1 at 80\% duty cycle}
        \label{fig:duty cycle plot3}
    \end{subfigure}
    \caption{Cumulative Distributions (CDF) of $90\%$ localization area of simulated BNS events obtained by different networks are shown. As mentioned in \ref{describing the duty cycles}, duty cycles for each of the L1, H1, V1, and K1 detectors are set to $80\%$. Three different cases considering A1 at duty cycles of $20\%$,  $50\%$, and $80\%$ are shown, along with taking into account two different noise sensitivities (A1 at O4 sensitivity and A1 at O5 sensitivity) and presented in Fig.~\ref{fig: duty cycle plot1}, \ref{fig:duty cycle plot2} and \ref{fig:duty cycle plot3} respectively. The blue solid lines represent the CDF obtained by the L1H1V1K1 network with the aforementioned duty cycle. The green solid lines represent the CDF constructed by the L1H1V1K1A1 network, where A1 is at \texttt{aLIGO} (O4) sensitivity. The red solid lines are the CDF constructed with L1H1V1K1A1 network, where A1 is at \texttt{A+ Design Sensitivity} (O5) sensitivity. In each of the subplots, the A1 detector is at a duty cycle, as mentioned in the captions of Fig.~\ref{fig: duty cycle plot1}, \ref{fig:duty cycle plot2} and \ref{fig:duty cycle plot3} respectively. The lighter curves represent the CDFs of $\Delta \Omega_{90\%}$ areas from different subnetworks over all possible $k$ for the corresponding GW network. The vertical black color dashed lines mark the median localization areas obtained by each network. The horizontal golden color dashed lines represent the fraction of events (in percentage) recovered under 10 sq. deg. by each detector network.}
    \label{fig:CDF plots O4, O5}
\end{figure*}
\begin{table*}[hbt]
\begin{ruledtabular}
\centering
\begin{tabular}{c c c c}
\multicolumn{1}{c}{Network} & \multicolumn{3}{c}{Fraction of events localized within $ < 10$ sq. deg.(in \%) \& Median $\Delta \Omega_{90\%}$ Area (sq. deg.)}\\
\hline
L1H1V1K1 &  & $59$\%  &  \\
(Median $\Delta \Omega_{90\%}$)  & & $6.6$  & \\
\cline{1-4}
&  A1 at $20\%$ duty cycle &A1 at $50\%$ duty cycle& A1 at $80\%$ duty cycle \\
\cline{2-4}

\hspace{4mm}L1H1V1K1+A1 (O4) &  64\% & 71\% & 77\%\\
(Median $\Delta \Omega_{90\%}$)  & 5.6 & 4.3 & 3.5 \\

\cline{1-4}
\hspace{4mm}L1H1V1K1+A1 (O5) &  66\% & 75\% & 84\%\\
(Median $\Delta \Omega_{90\%}$) & 4.9  & 3.4 & 2.4  \\
\end{tabular}
\end{ruledtabular}
\caption{We present the percentage of detected events localized within $10$ sq. deg sky area and the median $\Delta \Omega_{90\%}$ for different detector networks. The median $\Delta \Omega_{90\%}$ for detector networks are shown separately in the rows below the values representing the percentage of detected events localized within $10$ sq. deg sky area.  
The noise sensitivity of A1, along with the duty cycles associated with A1, are mentioned in the relevant table sections. Duty cycles for L1, H1, V1, and K1 detectors are fixed at $80\%$.}
\label{table 10 sq deg + median area}
\end{table*}

\section{Sky Localization Results}
\label{sky localization results}
In order to take into account the effect of duty cycles in the sky localization of our simulated BNS events, we first evaluate the probabilities associated with the effective duty cycles of each subnetwork of a detector network using Eq.~\eqref{equation_duty_cycle}. Each subnetwork is assigned a fixed number of events depending on their observation probabilities. Taking this into account and integrating all such cases across $3\leq k\leq N$ with area samples of corresponding events gives the localization $\Delta \Omega_{90\%}$ distribution related to the network duty cycle for a given GW network. 

The results of our simulations are presented in Fig.~\ref{fig:CDF plots O4, O5}. The Cumulative Distribution Function (CDF) plots in Fig.~\ref{fig:CDF plots O4, O5} are constructed from $\Delta \Omega_{90\%}$  sky area samples obtained by inverse sampling from the localization distributions for each subnetwork.
We find that with the L1H1V1K1 network, the median $90\%$ localization area $\Delta \Omega_{90\%}$ is $6.6$ sq. deg., meanwhile $59\%$ of the BNS sources are localized within less than $10$ sq. deg. area in the sky.
With the addition of an A1 detector to this network, we find significant improvements in the localization capabilities of the terrestrial detector network. It is also evident from Fig.~\ref{fig:CDF plots O4, O5} that duty cycles and detector noise sensitivities play a vital role in the effective localization of sources. We shall discuss these in detail as follows:\\

\subsection{A1 at \texttt{aLIGO}-O4 sensitivity}
\label{Explain CDF A1 at O4}
As a part of the five-detector network, the A1 detector is initially set to \texttt{aLIGO} O4 noise sensitivity. The median $\Delta \Omega_{90\%}$ area in the decreasing order are found to be $5.6$, $4.3$, and $3.5$  in sq. deg. when A1 is set to $20\%$, $50\%$, and $80\%$ duty cycle respectively. We find that $64\%$, $71\%$, and $77\%$ of the events are localized with less than $10$ sq. deg in sky area, given that A1 is at $20\%$, $50\%$, and $80\%$ duty cycles respectively. Our results suggest that even when A1 is at $20\%$ duty cycle, which can be interpreted as the early commissioning phase of the detector, the five-detector network reduces the median $\Delta \Omega_{90\%}$ localization uncertainty to $5.6$ sq. deg. in comparison to $6.6$ sq. deg. obtained by the four detectors L1H1V1K1 network. This reduction in the sky localization area plays a crucial role in the `tiled mode' search for EM counterparts undertaken by the EM facilities such as the GROWTH India Telescope~\cite{Kumar_2022} with a field of view of $~0.38$ sq. deg. in area, to tile the GW localization regions.
As the A1 detector is upgraded to be at $80\%$ duty cycle, the median localization area $\Delta \Omega_{90\%}$ remarkably reduces by approximately a factor of two in comparison to that achieved by the four detector L1H1V1K1 network. We reiterate that the L1, H1, V1, and K1 detectors are taken to be operating at $80\%$ duty cycles.\\

\subsection{A1 at \texttt{A+ Design} (O5) sensitivity}
\label{Explain CDF A1 at O5}
By upgrading the A1 configuration to \texttt{aLIGO A+ Design Sensitivity} (O5), the improvement in the localization capabilities of the five-detector network relative to the four-detector network as well as the five-detector network with A1 at O4 sensitivity is considerable. 
The median $\Delta \Omega_{90\%}$ localization uncertainties in the decreasing order are found to be $4.9$, $3.4$, and $2.4$ sq. deg. in area, when A1 is set to $20\%$, $50\%$ and $80\%$ duty cycle respectively. We find that $66\%$, $75\%$, and $84\%$ of the events are localized with less than $10$ sq. deg in sky area, given that A1 is at $20\%$, $50\%$ and $80\%$ duty cycles respectively, where A1 is set at \texttt{A+} sensitivity. We observe that with the A1 detector operating at $50\%$ duty cycle, the median localization area $\Delta \Omega_{90\%}$ reduces by a factor of two with respect to the values obtained by the L1H1V1K1 network. As A1 reaches its target operating point with $80\%$ duty cycle, we find the median $\Delta \Omega_{90\%}$ to reduce by a factor of three against the median localization achieved with the four-detector network. As mentioned previously, for A1 operating at O4 sensitivity and $80\%$ duty cycle, the median $\Delta\Omega_{90\%}$ is $3.5$ sq. deg., whereas this reduces to $2.4$ sq. deg. when A1 is set to $80\%$ duty cycle and at O5 sensitivity.

\begin{figure*}
    \centering
    \includegraphics[width=0.9\textwidth, height=0.55\textwidth, clip=True]{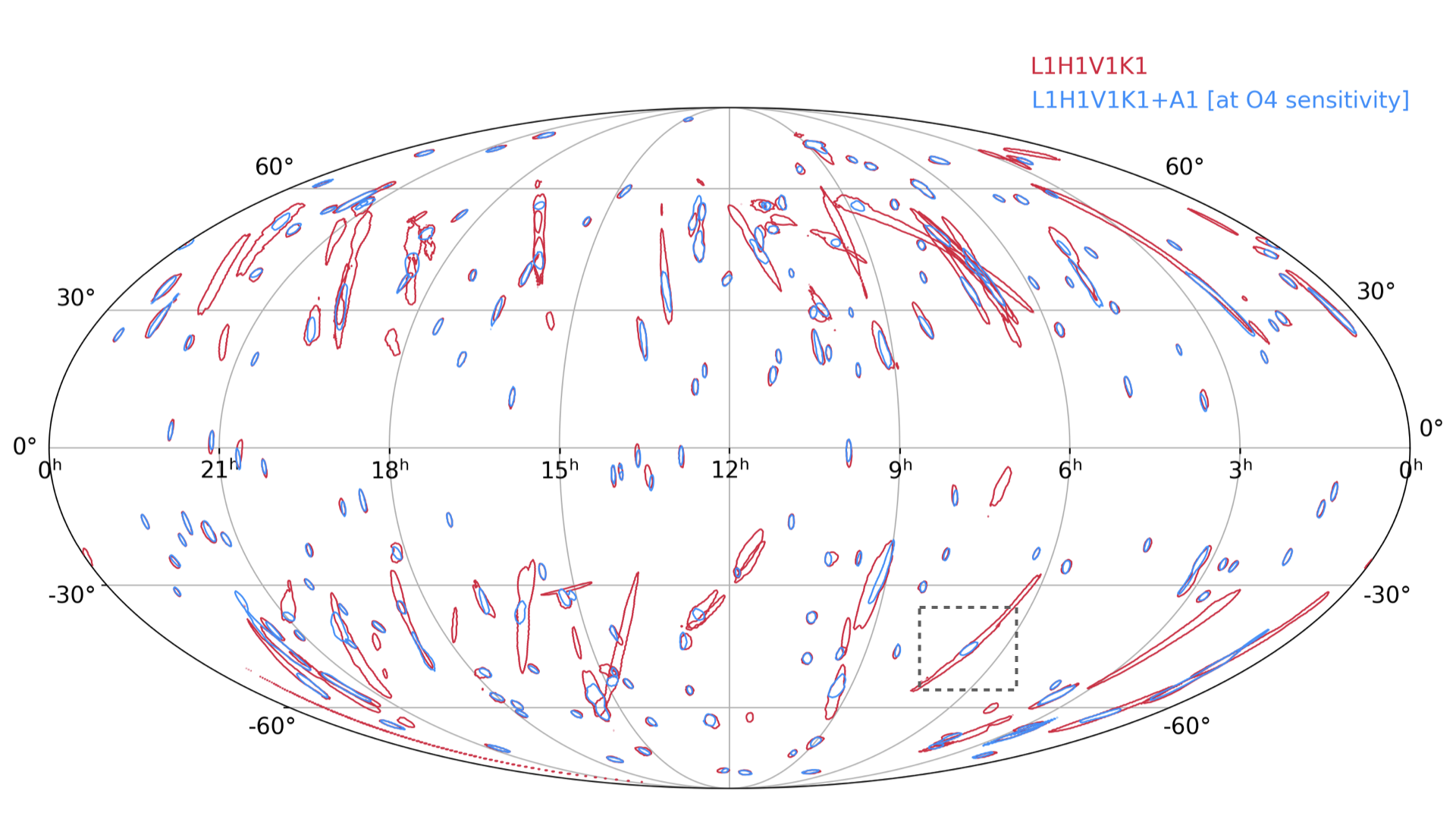}
    \caption{The crimson and blue probability contours correspond to 90\% credible sky area ($\Delta \Omega_{90\%}$) localization of the events by L1H1V1K1 and L1H1V1K1A1 networks, respectively 
    for 191 events out of our 500 simulated BNS events. Here, A1 is at O4 sensitivity, and the events are subthreshold in A1. The contours marked in a dashed square box are discussed in Section~\ref{Subthreshold Events in LIGO-Aundha}. This skymap plot exhibits the improvement achieved in the localization of BNS events by the inclusion of an A1 detector, even when it does not contribute to the detection.} 
    \label{fig: A1 subthreshold O4 skymap}
\end{figure*}

The A1 detector, with an upgraded O5 sensitivity, leads to a two-fold impact on the network capabilities. On the one hand, it leads to an increase in the number of detections (events satisfying the SNR threshold $\rho_{\text{th}} \geq 6 $) in the subnetworks, including A1. It also results in an increase in network SNR, which is one of the important factors contributing to the effective localization of sources. By adding A1 to the GW network, there is an increase in the observation probability of three or more detectors by $7\%$ (for A1 operating at the early $50\%$ duty cycle) in comparison to the four detector networks. We summarize a few important remarks about the localization $\Delta \Omega_{90\%}$ for the different network configurations in Table \ref{table 10 sq deg + median area}. Note that the CDF plots in Fig.~\ref{fig:CDF plots O4, O5} may indicate slightly lower median $\Delta \Omega_{90\%}$ localization values than those obtained in related studies~\cite{Pankow_2020, Pankow_2018}. One of the reasons being that the events are analyzed from $f_{\text{low}} = 10 \hspace{0.5mm}\Hz$, which increases the effective bandwidths, as well as due to the exclusion of cases with two detectors subnetworks or single detectors participation in source localization.
 
The focus of this study is to explore the localization capabilities of the GW network with A1 with possibilities leading to potential EM follow-ups as well as providing a better ground for astrophysical and cosmological investigations.

\section{Localization of simulated BNS events subthreshold in A1}
\label{Subthreshold Events in LIGO-Aundha}
During the observation of GW170817, the event was detected in L1 and H1 detectors but was below the detection threshold in the V1 detector. Yet, the presence of V1 contributed to localizing the source to a few tens of sq. deg. Out of the $500$ simulated BNS events described in our previous discussion, a total of $191$ events detected in the five-detector network were found to be subthreshold ($\rho_{\text{th}}<6$) in A1 is at \texttt{aLIGO} O4 sensitivity. We compare the sky-localizations for these events obtained from a four-detector L1H1V1K1 network to those achieved by the L1H1V1K1A1 network. Since these events are subthreshold in A1, the contribution of A1 in improving the network SNR is negligible. 
Yet, the presence of an A1 detector leads to an improvement in reducing the localization uncertainties of these events. This is shown in Fig.~\ref{fig: A1 subthreshold O4 skymap}. 
We find that even in the case where these events are subthreshold in A1 (at \texttt{aLIGO} O4 sensitivity), the percentage of events localized with less than $10$ sq. deg. in the sky increase from $72\%$ to $89\%$ in comparison to the L1H1V1K1 network. Even though the CDFs (for $\Delta \Omega_{90\%}$) used for the estimation of these improvements are not very smooth due to a lesser number of such events (191 in this case), but nevertheless they summarize the essence of overall nature of the improvement well enough.

As the noise sensitivity configuration of A1 is upgraded to \texttt{aLIGO A+ Design Sensitivity}, there is an increase in the number of detections in the A1 detector.
In this case, the number of events that are subthreshold in the A1 detector reduces from 191 to just 44 out of all the 500 simulated BNS sources. Due to the improved sensitivity, further improvements in the localization of such events are achieved in comparison to the localizations obtained relative to the four-detector L1H1V1K1 network and L1H1V1K1A1 network with A1 at O4 sensitivity. Note that in the context of this section, we do not consider the duty cycles for these networks. Therefore, we make a direct comparison between the localization results achieved for such events with the L1H1V1K1 network and the L1H1V1K1A1 network. For instance, the event marked in Fig.~\ref{fig: A1 subthreshold O4 skymap} is localized to $44$ sq. deg. with L1H1V1K1 network. The same event when detected by the L1H1V1K1A1 network with A1 is at \texttt{aLIGO} O4 sensitivity, is recorded at an optimal SNR value $\rho_{A1} = 3.1$ in A1 detector ($\rho_{A1} < \rho_{\text{th}}$) and is localized to $\sim 6$ sq. deg. Meanwhile, when A1 is set to \texttt{aLIGO A+ Design Sensitivity}, this event is localized to $\sim 3.5$ sq. deg. area in the sky. The baselines added to the network with the addition of the A1 detector, and its antenna patterns are some of the factors leading to better localization of such events. An improved noise PSD results in an increase in effective bandwidth and hence leads to the reduction in localization uncertainties.

\section{Experiments with GWTC-like events in real noise}
\label{GWTC-Events}
In the preceding section, we showed that even if a detector does not detect an event, it nevertheless adds a valuable contribution to the network in localizing the source. In this section, we provide an illustration of how the incorporation of an additional detector could have facilitated the source localization of events from GWTC for compact binary mergers. The two BNS events, GW170817 and GW190425, and a neutron star-black hole (NSBH) event, GW200115, are chosen as examples from GWTC for this purpose. In our analysis, we consider A1 as a supplementary detector. We simulate the aforementioned events and inject them into real detector noise to account for a realistic scenario. The noise strains from the L1, H1, and V1 detectors were acquired by using the \texttt{GWpy}~\cite{gwpy} python package, which allows the extraction of noise strain timeseries from the datasets publicly available on GWOSC~\cite{2022AAS...24034809B}. The noise strain for A1 is taken to be that of the detector, which recorded the least SNR during the observation of these events. In fact, among all the detectors observing these events, the lowest individual SNR was recorded in the Virgo detector. The noise strain data for all the detectors is chosen hundreds of seconds away from the trigger times of the GW events in consideration. Even though the A1 noise strain is taken from V1 data, the noise strain data for both detectors belong to different stretches of data. The PE analysis for these events is performed from a lower seismic frequency of $20$ Hz.

\subsection{Noise Strain and PSDs} 
A noise strain data of a fixed segment length (360 seconds for GW170817 and GW190425; 64 seconds for GW200115) is used in our analysis, which is cleaned by a high-pass filter of $4$th order and setting the frequency cut-off at $18$ Hz. We analyze the event from a lower seismic cutoff frequency of $20$ Hz, which is illustrative of the observing runs associated with their detections. The estimation of noise PSD uses $2$ seconds overlapping segments of the strain data with the implementation of the median-mean PSD estimation method from \textsc{PyCBC}~\cite{usman2016pycbc}. The noise PSD for A1 is constructed from the strain data of the V1 detector.
    
\subsection{Choice of injection parameters:}
    \subsubsection{GW170817-like event} 
        The intrinsic detector-frame parameters $(m_1^{\text{det}}, m_2^{\text{det}}, \chi_{1z}, \chi_{2z})$ and extrinsic parameters like inclination ($\iota$) take values chosen by evaluating the MAP values from the posterior samples of detector-frame parameters in the \texttt{Original BILBY results file}~\cite{Romero_Shaw_2020} for GW170817. We assume the BNS system with spins aligned in the direction of orbital angular momentum. The sky location coordinates of NGC $4993$-the potential host galaxy of the GW170817 event, are taken as the injection values for $(\alpha, \delta)$ sky position parameters~\cite{Ebrova:2018gtz}. The luminosity distance takes the value $d_L = 40.4$ Mpc~\cite{Hjorth_2017} for our simulated BNS system. The polarization angle is taken to be zero ($\psi=0$) since the associated posterior samples in  \texttt{Original BILBY results file} are found to be degenerate. As mentioned previously, the simulated signal is injected in uncorrelated real noise in the detectors. The simulated signal is generated using the \texttt{IMRPhenomD} waveform model. The source parameters are recovered using the \texttt{TaylorF2} waveform model.
        
    \subsubsection{GW190425-like event} 
        The intrinsic (detector-frame) and extrinsic parameter values are chosen by evaluating the MAP values
        of the posterior samples for parameters obtained from 
        \texttt{C01:IMRPhenomPv2\_NRTidal:LowSpin} LIGO analysis file of GW190425 event~\cite{ligo_scientific_collaboration_and_virgo_2022_6513631}. The polarization angle is chosen to be $\psi=0$ as the posterior samples for $\psi$  from the LIGO analysis follow a uniform distribution. We generate the simulated signal using the \texttt{IMRPhenomD} waveform model. The source parameters are recovered using the \texttt{TaylorF2} waveform model for the PE analysis.

    \subsubsection{GW200115-like event} 
        For simulating the NSBH event, we choose the intrinsic (detector-frame) and extrinsic parameter values by evaluating the MAP values of the posterior samples for parameters from \texttt{C01:IMRPhenomNSBH:LowSpin} file of GW200115 LIGO analysis~\cite{ligo_scientific_collaboration_and_virgo_2021_5546663}. We take the polarization angle $\psi=0$. We generate the simulated signal using the \texttt{IMRPhenomD} waveform model. For recovering the source parameters during the PE analysis, we again use the \texttt{IMRPhenomD} waveform model, as it also accounts for the post-inspiral regime, which occurs within the LIGO-Virgo band for the NSBH system.

\subsection{Analysis and Configurations} 
The prior distributions and prior boundaries for ($\mathcal{M_{\text{det}}}, q, \chi_{1z}, \chi_{2z}, V_{\text{com}} $) parameters, chosen for the three simulated events are presented in Table \ref{Hybrid table for GWTC events}. The priors for the parameters ($t_c$, $\alpha$, $\delta$, $\iota$, $\psi$)  are same as that shown in Table \ref{tab:priordistr} and hence are not seperately mentioned here. The Bayesian PE analysis follows a similar methodology of generating interpolants for the likelihood function, as discussed previously. The analysis involves the generation of RBF nodes.  We specify the total number of RBF nodes ($N_{\text{nodes}}$) by mentioning the number of nodes sampled from a multivariate Gaussian $\mathcal{N}(\vec \lambda^{\text{cent}}, \mathbf{\Sigma})$, represented by $N_\text{Gauss}$; meanwhile the number of nodes uniformly sampled around $\vec\lambda^{\text{cent}}$ are represented as $N_\text{Unif}$ for each event. The \texttt{dynesty} sampler configurations are also mentioned in Table \ref{Hybrid table for GWTC events} for the three simulated events. 

\begin{table*}[hbt]
\def\arraystretch{2}
\begin{ruledtabular}
\centering
\begin{tabular}{l c c c c l}

\multirow{1}{*}{}&\multicolumn{1}{c|}{GW170817-like} & \multicolumn{1}{c|}{\centering \hspace{-2mm}GW190425-like}& \multicolumn{1}{c}{\hspace{-5mm}GW200115-like}& \multirow{1}{*}{} \\
\cline{2-4}
Parameter & Prior Range & Prior Range & Prior Range & Prior Distribution\\
\hline
$\mathcal{M_{\text{det}}}$  & 
$[\mathcal{M}^{\text{cent}}_{\text{det}} \pm 0.0002]$& $[\mathcal{M}^{\text{cent}}_{\text{det}} \pm 0.0005]$& $[\mathcal{M}^{\text{cent}}_{\text{det}} \pm 0.0011]$ & 
$\propto \mathcal{M_{\text{det}}}$ \\

$q$  & $[1, 1.14]$ & $[1, 1.28]$  & $[q^{\text{cent}} \pm 0.1]$ & $\propto \left [ (1 + q)/q^3 \right ]^{2/5}$  \\

$\chi_{1z, 2z}$ & 
$[\chi_{1z, 2z}^{\text{cent}} \pm 0.0025]$   & 
$[\chi_{1z, 2z}^{\text{cent}} \pm 0.0025]$   &
$[\chi_{1z, 2z}^{\text{cent}} \pm 0.0025]$   &
Uniform  \\

$V_{\text{com}}$ & $[5e3, 1e8]$ & $[5e3, 4e8]$ & $[1e6, 3e8]$ & Uniform \\
\hline

\multirow{2}{*}{No. of RBF Nodes} & $N_\text{nodes}=800$ & $N_\text{nodes}=800$ & $N_\text{nodes}=1100$ & \\

& ($N_\text{Gauss}=20\%$, $N_\text{Unif}=80\%$) & ($N_\text{Unif}= 100\%$) & ($N_\text{Gauss} = 10\%$, $N_\text{Unif}= 90\%$) & \\

\hline

RBF Parameters & $\epsilon=10, \nu=7$ & $\epsilon=10, \nu=7$ & $\epsilon=30, \nu=10$ & $\phi = \exp(-(\epsilon r)^2)$ \\

\hline

\multirow{2}{*}{Sampler Configurations} & \texttt{nLive}=500 & \texttt{nLive}=1500 & \texttt{nLive}=500 & \\

\addlinespace[-5mm] 

& \texttt{nwalks}=100 & \texttt{nwalks}=500 & \texttt{nwalks}=100 & \\

\addlinespace[-5mm]

& \texttt{sample}=``rwalk" & \texttt{sample=}=``rwalk" & \texttt{sample}=``rwalk" & \\

\addlinespace[-5mm] 

& \texttt{dlogz} $=0.1$ & \texttt{dlogz} $=0.1$ & \texttt{dlogz} $=0.1$ & \\

\end{tabular}
\end{ruledtabular}
\caption{The prior parameter space is presented in the top part of the table. The hybrid placement for the RBF nodes is mentioned by specifying $N_\text{Gauss}$ and $ N_\text{Unif}$ for each event, followed by the Gaussian RBF kernel parameters. The \texttt{dynesty} sampler configurations chosen for the PE analysis of each event are also presented in the end. The priors for the extrinsic source parameters ($t_c$, $\alpha$, $\delta$, $\iota$, $\psi$) are same as that shown in Table \ref{tab:priordistr} and hence are not mentioned here.}
\label{Hybrid table for GWTC events}
\end{table*}

\begin{figure}[hbtp] 
    \centering
    \hspace{-13mm}
    \includegraphics[width=0.4\textwidth, height=0.4\textwidth, clip=True]{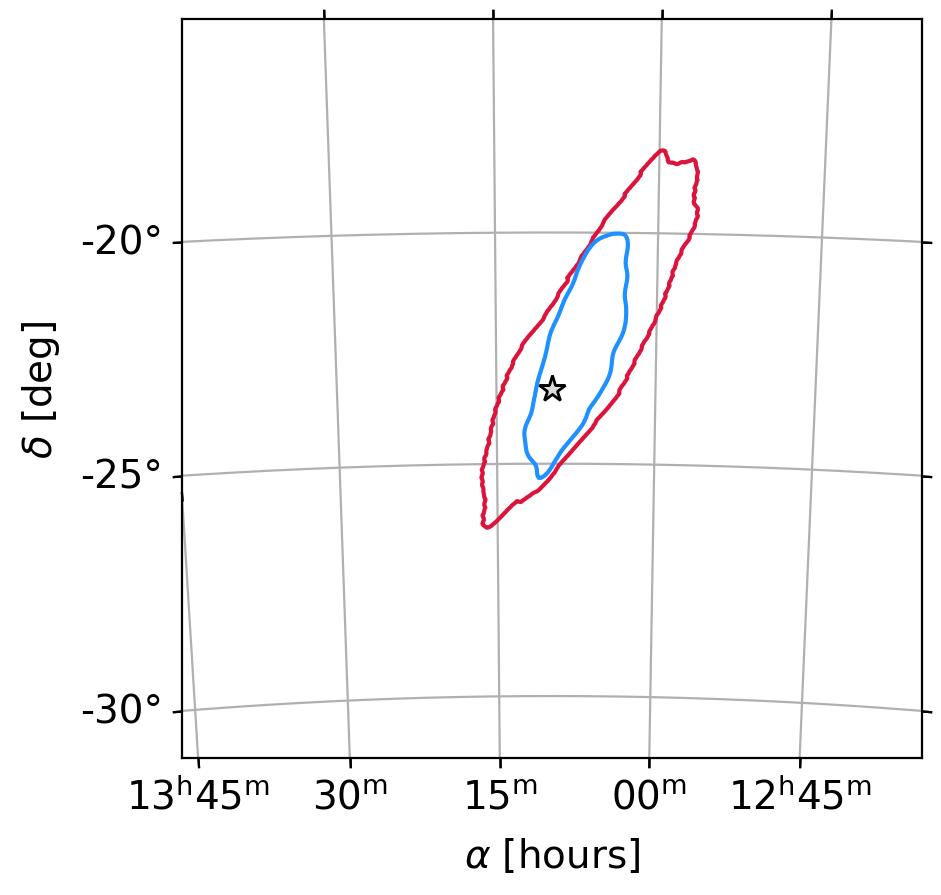}
    \caption{The crimson and blue probability contours represent $90$\% credible localization area ($\Delta \Omega_{90\%}$) for the GW170817-like event by L1H1V1 and L1H1V1A1 networks respectively. Noisy real strain data from V1 are used as a surrogate for A1 in this exercise. The star marks the true location $(\alpha, \delta)$ for the source. Though A1 does not contribute to the network SNR, its presence leads to a significant improvement in source localization.}
    \label{fig: GW170817 with A1}
\end{figure}

The network comprising the L1, H1, and V1 detectors detected the GW170817 event. As discussed earlier, we simulate a GW170817-like signal in the non-Gaussian real noise and find the source localization uncertainty
in the presence of an A1 detector added to the L1H1V1 network. The matched-filter SNR in L1, H1, V1 and A1 are $22.6$, $18.6$, $5.4$ and $6.3$ respectively for the given noise realization. Even though in this case, the addition of A1 to the L1H1V1 network does not lead to any considerable improvements in the network matched-filter SNR for the event, yet a significant reduction in $90\%$ credible localization area is observed. We find the localization uncertainty $\Delta \Omega_{90\%}$ to be $15$ sq. deg for the L1H1V1 network, whereas the localization area $\Delta \Omega_{90\%}$ reduces to $6$ sq. deg. with L1H1V1A1 network. Hence, the localization uncertainty is reduced by a factor of more than two in this case. The localization probability contours representing $\Delta \Omega_{90\%}$ obtained from the two different networks for GW170817-like event is presented in Fig.~\ref{fig: GW170817 with A1}. For a GW190425-like event, we compare the sky localization with the then-observing network of the L1V1 network to the L1V1A1 network. The sky localization uncertainty ($\Delta \Omega_{90\%}$) reduces from 9350 sq. deg. with the L1V1 network to a sky region of area 212 sq. deg with the L1V1A1 network. The matched filter SNR in L1, V1, and A1 are $10.1$, $5.1$, and $5.3$, respectively, for this case. It is evident that the event in V1 and A1 is at subthreshold SNR for the given noise realization. Yet, there is a contribution in reducing the sky localization areas. For the case of the GW200115-like event, the source localization with the L1H1V1 network, which was the observing network during the real event, is compared to that with the L1H1V1A1 network. The source localization error ($\Delta \Omega_{90\%}$) is reduced from $662$ sq. deg. obtained with L1H1V1 to $87$ sq. deg. achieved with L1H1V1A1. Here, we observe that the majority of the SNR is accumulated by the initial two LIGO detectors. Meanwhile, V1 and A1 contribute negligibly to improving the network SNR. This is because both V1 and A1 are at similar noise sensitivities for the aforementioned events. 

Note that these results vary with different realizations of the detector noise. Nevertheless, the antenna patterns and baselines added to a network by incorporating an additional detector (here, A1) may lead to an enhancement in the localization abilities of the network, even if the signal is subthreshold in one of the detectors.

\subsection{Degeneracy between luminosity distance and inclination angle}
\label{dL_iota_degeneracy_sec}
The GW190425-like event, when observed by the L1V1 network, shows a degeneracy between luminosity distance and inclination angle parameters, which was also observed for the real event. 

\begin{figure}[htbp]
    \centering
    \includegraphics[width=0.52\textwidth, height=0.55\textwidth, clip=True]{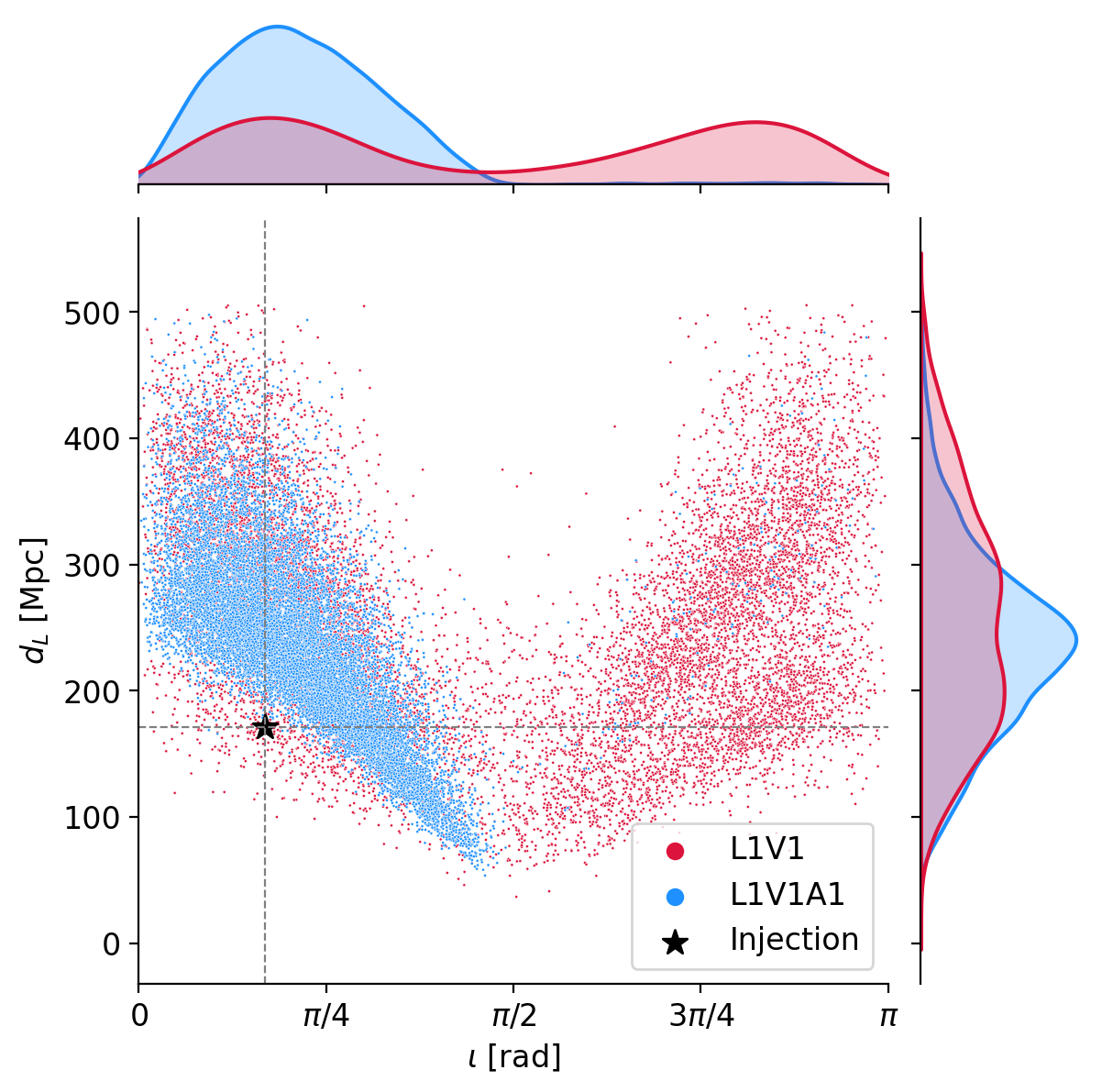}
    \caption{The crimson and blue points mark the posterior samples over $d_L$ and $\iota$ for the two-detector L1V1 and three-detector L1V1A1 networks respectively. The posteriors are constructed over 28 different noise realizations. This signifies that a three-detector network (L1V1A1) with a third detector contributing negligibly to the network SNR (in this case, A1) is able to play an important role in resolving the $\iota$--$d_L$ degeneracy relative to a two detector network (L1V1).}
    \label{fig:dL_vs_iota}
\end{figure}

As the number of detectors in the network increases from L1V1 to L1V1A1, we observe a resolution of the distance-inclination angle degeneracy. For further investigation, we present the case for GW190425-like events with $28$ different real non-Gaussian noise realizations. The noise strains and PSD are obtained as mentioned at the beginning of Section~\ref{GWTC-Events}, where different noise strains correspond to different segments of detector strain data. The events for which the injected chirp-mass ($\mathcal{M}_c$) is within the $90\%$ credible interval of the posterior samples are chosen.
 
The results are summarized in Fig.~\ref{fig:dL_vs_iota}. The degeneracy between the luminosity distance ($d_L$) and inclination angle ($\iota$) parameters is resolved with an additional detector (here A1), even when the contribution of A1 in increasing the network SNR is not appreciable relative to the two detector network (L1V1).
Note that, here, GW190425-like events are generated with a waveform model (\texttt{IMRPhenomD}), which does not include higher-order modes. Also, both the compact objects (in this case: BNS) are of approximately equal masses i.e. $q\approx 1$. Hence, we can safely assume that the higher-order modes do not play a role in the resolution of degeneracy between the parameters. We obtain similar results on relaxing the condition over $\mathcal{M}$ and performing a similar analysis for $60$ different noise realizations. An investigation addressing the luminosity distance and inclination angle degeneracy for BNS systems has also been done in~\cite{Rodriguez_2014}. It is not clear that a better measurement of both the polarizations ($h_+$ \& $h_{\times}$) in a larger network leads to a more precise measurement of the inclination, especially for face-on systems ($\iota< 45$ deg.). In our study, we show the result as an empirical observation for a GW190425-like event. An extensive study constraining the inclination angle with a network of GW detectors has been performed by Usman \textit{et al.}~\cite{Usman_2019}.

The improvement in the measurements of luminosity distance has direct implications in cosmology, as mentioned in Section~\ref{sec:intro}. The accurate measurements of inclination angle may lead to improvements in the constraints on the models for gamma-ray bursts and X-ray emissions from BNS mergers~\cite{Finstad_2018}. Similar improvements in the measurements of luminosity distance and inclination angles for binary black hole mergers by a three-detector network relative to a two-detector network have been obtained in~\cite{Saleem_2021}. 

\section{Conclusion}
\label{Conclusion_section}
The addition of A1 to the GW network is observed to improve the overall localization capabilities of the global detector network, even when A1 is in its early commissioning stages. To estimate the source parameters, we performed a full Bayesian PE from a lower cut-off frequency  $f_{\text{low}}= 10$ Hz, which is representative of the future LVK Collaboration analysis of GW sources. We find that addition of A1 detector (at \texttt{aLIGO} O4 sensitivity) to the GW network leads to a reduction of the median $\Delta \Omega_{90\%}$ area to $5.6$, $4.3$, and $3.5$ sq. deg. for cases where A1 is operating at $20\%$, $50\%$, and $80\%$ duty cycles respectively, in comparison to the median $\Delta \Omega_{90\%}$ area of $6.6$ sq. deg. obtained with the four detector L1H1V1K1 network for BNS sources with potential for multi-messenger follow-ups.

Our results suggest that an expanded GW detector with at an early phase A1 operating at a $20\%$ duty cycle and operating at a weaker sensitivity (\texttt{aLIGO} O4) as compared to the other LIGO detectors (\texttt{aLIGO} A+ Design Sensitivity) is capable of localizing $64\%$ of these BNS sources under $10$ sq. deg in comparison to $56\%$ by the four detector network. With the imminent improvement in the duty cycle and noise PSD of the A1 detector, an apparent enhancement in the localization capabilities of the GW network is observed (Refer Table~\ref{table 10 sq deg + median area}). With the addition of an A1 detector to the GW network, the observation probability for the sub-networks of $k\geq3$ detectors increases, leading to a decrease in localization uncertainties in the sky area. This allows for an optimized ``tiled mode'' search for post-merger emissions by telescopes such as the GROWTH India facility with a field of view of the order of $\sim0.4$ sq. deg. in sky area. We show that improvements in duty cycles and noise sensitivity for A1 detector play a crucial role in enhancing the localization capabilities of the GW network. Hence, in order to get the maximal payoff from the addition of the A1 detector, efforts should be made towards maximizing the operational duty cycle and improving the noise sensitivity as soon as the detector becomes operational. 

Furthermore, we show that even for BNS sources that are sub-threshold in A1, the sky-localization uncertainties with the five-detector L1H1V1K1A1 network are reduced in comparison to that obtained from the four-detector L1H1V1K1 network. Thus, even in a situation where A1 does not detect the BNS event independently, it plays a crucial role in pinpointing the sources that enable a fast and efficient electromagnetic follow-up by ground and space-based telescopes.

Taking the examples of two BNS events and one NSBH event from GWTC, we show the possible source localization improvements with A1 as an additional detector in the network with real noise. For this exercise, the real noisy strain data from Virgo is used as surrogate noise in A1 detector to simulate a scenario where the Indian detector is observing the event but has not achieved its design sensitivity. 

We reaffirm the role of an additional detector (A1 in our case) in resolving the degeneracy between luminosity distance and inclination angle parameters relative to a two-detector network for a GW190425-like BNS source. This is shown by reconstructing the source parameters for GW190425-like BNS events in real, non-Gaussian noise, with the L1V1 and L1V1A1 detector networks, respectively, where data samples from V1 are used as surrogates for A1. 

\section{Discussion}
\label{Discussion}
In order to maximize the incentives from the GW detection of BNS sources, the EM follow-up of these events is of utmost importance. A1, joining the network of terrestrial GW detectors in the early 2030s, will enhance the localization capabilities of the network. We studied the impact of the addition of A1 in the detector network in the localization of BNS sources with moderately high signal-to-noise ratios. 

The observation of an event with three or more detectors working in conjunction is fundamental for achieving localization uncertainties small enough so as to allocate telescope time for subsequent electromagnetic follow-ups. Our results presented in Fig.~\ref{fig:CDF plots O4, O5} from Section~\ref{sky localization results} can be considered optimistic, owing to the assumption of a BNS event being observed by more than two detectors at any given time. 
Including sub-networks of two detectors will lead to the broadening of the distribution of localization uncertainties, causing a slight shift to the right in the cumulative distribution functions (CDFs) shown in Fig.~\ref{fig:CDF plots O4, O5}.
However, this is beyond the scope of this work, and a more realistic study taking the case of two detector subnetworks into account can be performed in the future. Along with this, considering the case where one or more detectors turn out to be at duty cycles that are lower than expected, as is the case for Virgo and KAGRA during the O4 run, can provide a more realistic account depicting the localization capabilities of a GW network. For instance, in the context of our study, the median $\Delta \Omega_{90\%}$ area of $\sim 13.5$ sq. deg. is obtained with the four detector L1H1V1K1 network, where L1 and H1 are at $80\%$ duty cycle and both V1 and K1 detectors are operating at a lower $20\%$ duty cycle. With the addition of A1 detector to this network, where A1 is set to \texttt{aLIGO} O4 sensitivity and operates at $20\%$ duty cycle (same as that of V1 and K1), there is a significant reduction in median $\Delta \Omega_{90\%}$ area to $\sim 8$ sq. deg.
A case study of the localization capabilities considering only the three LIGO detectors (L1, H1, and A1) is presented in ~\cite{Saleem_2021}, where all three are considered to be at A+ sensitivity. 

For this study, we generate the BNS events in Section~\ref{motivating for SNR range choice & injection parameters} using the \texttt{IMRPhenomD} waveform model and reconstruct the source parameters using the \texttt{TaylorF2} model template waveforms. Using a waveform model that includes tidal deformability parameters, higher-order modes, and other physical effects captured by additional intrinsic parameters in the analysis can make the study more comprehensive. We aim to 
incorporate the tidal parameters and higher-order modes within the meshfree framework in the future. This extension will enable us to achieve a more comprehensive and rigorous analysis. We have also fixed the values of $\iota$ and $\psi$ as shown in Section~\ref{motivating for SNR range choice & injection parameters}. This may also have an effect on the localization results. For a more general treatment, the events under consideration should be generated such that all the parameters should be allowed to vary in parameter space. This shall allow for a more exhaustive assessment of the localization capabilities of different GW networks. We simulate uncorrelated Gaussian noise in the detectors for our analysis. In this context, it has been shown by Berry \textit{et al.}~\cite{Berry_2015} that no appreciable impact is observed in the localization results for the case of simulated signal injected in real detector noise. 

Another aspect that might affect the sky localization area evaluated from the Bayesian posterior samples is the narrow prior boundaries taken over the intrinsic parameters. For consistency, we evaluate the sky localization areas from the posterior samples with wide boundaries over intrinsic parameters using another rapid PE method (relative binning in this case) and compared the results with the meshfree framework adopted here. We find that the difference between the localization areas obtained from these two approaches is not significant enough to affect the localization results, at least for a network involving three or more detectors. The results showcased in this work serve as a demonstration of what can be accomplished by adding A1 as a new detector to the GW network. The primary emphasis is on evaluating the GW network's ability to pinpoint the source of GW, particularly in the context of potential electromagnetic follow-up observations.

\begin{acknowledgements}
We would like to thank Varun Bhalerao, Gaurav Waratkar, Aditya Vijaykumar, Sanjit Mitra, and Abhishek Sharma for useful suggestions and comments. We especially thank the anonymous referee for their careful review and helpful suggestions.

S.~S. is supported by IIT Gandhinagar. L.~P. is supported by the Research Scholarship Program of Tata Consultancy Services (TCS). A.~S. gratefully acknowledges the generous grant provided by the Department of Science and Technology, India, through the DST-ICPS cluster project funding. We thank the HPC support staff at IIT Gandhinagar for their help and cooperation. The authors are grateful for the computational resources provided by the LIGO Laboratory and supported by the National Science Foundation Grants No. PHY-0757058 and No. PHY-0823459. This material is based upon work supported by NSF's LIGO Laboratory, which is a major facility fully funded by the National Science Foundation. 

This research has made use of data or software obtained from the Gravitational Wave Open Science Center~\cite{gwosc_web}, a service of the LIGO Scientific Collaboration, the Virgo Collaboration, and KAGRA. This material is based upon work supported by NSF's LIGO Laboratory, which is a major facility fully funded by the National Science Foundation, as well as the Science and Technology Facilities Council (STFC) of the United Kingdom, the Max-Planck-Society (MPS), and the State of Niedersachsen/Germany for support of the construction of Advanced LIGO and construction and operation of the GEO600 detector. Additional support for Advanced LIGO was provided by the Australian Research Council. Virgo is funded through the European Gravitational Observatory (EGO), the French Centre National de Recherche Scientifique (CNRS), the Italian Istituto Nazionale di Fisica Nucleare (INFN), and the Dutch Nikhef, with contributions by institutions from Belgium, Germany, Greece, Hungary, Ireland, Japan, Monaco, Poland, Portugal, Spain. KAGRA is supported by the Ministry of Education, Culture, Sports, Science and Technology (MEXT), Japan Society for the Promotion of Science (JSPS) in Japan; National Research Foundation (NRF) and Ministry of Science and ICT (MSIT) in Korea; Academia Sinica (AS) and National Science and Technology Council (NSTC) in Taiwan.
\end{acknowledgements}

\bibliography{references}

\end{document}